\documentclass[a4paper,11pt]{article}
\usepackage{float}
\usepackage[utf8]{inputenc}
\usepackage[T1]{fontenc}
\usepackage[]{units}
\usepackage[margin = 2.5cm]{geometry}
\usepackage{mathrsfs}
\usepackage{subfigure}
\usepackage{multirow}
\usepackage{multicol}
\usepackage{colortbl}
\usepackage{subcaption}
\usepackage{comment}
\usepackage{amsmath}
\usepackage{extarrows}
\usepackage{mathtools}
\usepackage{amsfonts}
\usepackage{amssymb}
\usepackage{enumerate}
\usepackage{diagbox}
\usepackage{amsthm}
\usepackage{bm}
\usepackage{textgreek}
\usepackage{tabularx}
\usepackage{float}
\usepackage{threeparttable}
\usepackage{graphicx}
\usepackage{textcomp} 
\usepackage{epstopdf}
\usepackage{stackengine}
\usepackage{enumitem}
\usepackage{cancel}
\usepackage{hyperref}
\usepackage{tikz}
\usetikzlibrary{shapes.geometric, arrows, positioning, fit, backgrounds}
\newcommand{\Var}{\operatorname{Var}}

\usepackage[svgnames]{xcolor}
\usepackage[labelfont=bf]{caption}
\usepackage{subcaption}
\usepackage{amsmath}
\usepackage{blindtext}
\usepackage{tikz}

\usepackage{pdflscape}
\usepackage{geometry}
\usepackage{rotating}
\usepackage{multirow}
\usepackage{blindtext}
\usepackage{wrapfig}
\usepackage{rotating}
\usepackage{booktabs}
\usepackage{caption}
\usepackage{subcaption}
\usepackage{booktabs}
\usepackage{xltabular}
\usepackage{titling}
\predate{}
\postdate{}
\usepackage{authblk}
\usepackage{booktabs, makecell, longtable}
\usepackage{pgfkeys}
\usepackage{tikz}
\usepackage{tikz}
\usetikzlibrary{shapes.geometric,arrows.meta,positioning}
\usepackage{multirow}
\usepackage{lineno} 

\usepackage{amsthm} 
\usepackage{float}

\hypersetup{
    colorlinks=true,
    linktocpage=true,
    pdfstartpage=1,
    pdfstartview=FitV,
    breaklinks=true,
    pdfpagemode=UseNone,
    pageanchor=true,
    pdfpagemode=UseOutlines,
    plainpages=false,
    bookmarksnumbered,
    bookmarksopen=true,
    bookmarksopenlevel=1,
    hypertexnames=true,
    pdfhighlight=/O,
    linkcolor=red,  
    citecolor=blue,
    urlcolor=blue,
    pdfsubject={},
    pdfkeywords={},
    pdfcreator={pdfLaTeX}
}

\usepackage{natbib}

\usepackage{epstopdf}
\AppendGraphicsExtensions{.tiff}

\usepackage{algorithm}
\usepackage{algpseudocode}
\usepackage{adjustbox}

\title{\textbf{ A spatial duration-augmented framework for drought persistence 
}}
\author[1]{Touqeer Ahmad}
 \author[1]{Thordis Linda Thorarinsdottir} 
\affil[1]{\textit{Department of Mathematics, University of Oslo, P.O. Box 1053 Blindern, 0316 Oslo, Norway.}}

\date{} 
\begin{document}
\maketitle
\begin{abstract}

\noindent
Drought monitoring across heterogeneous climate regions requires explicit modeling of drought persistence rather than relying solely on index classification. We develop a Bayesian spatial, duration-augmented framework for drought persistence that captures both temporal memory and spatial dependence. Because persistence estimates depend on the underlying drought index, we also assess the influence of (i) reference evapotranspiration (RET) formulations and (ii) the choice of probability distribution used to standardize the climatic water balance. The framework is applied to a high-resolution (0.25$^\circ \times$ 0.25$^\circ$) dataset over 730 grid cells across Italy.
The choice of RET formulation (Penman-Monteith, Hargraves, Thornthwaite) has only a marginal influence on the overall drought signal: spatially averaged SPEI series at the 3-, 6-, and 12-month scales are nearly identical across methods. By contrast, the proposed bulk-and-tails (BATs) distribution substantially improves index estimation, consistently outperforming commonly used alternative models across all RET methods and timescales and achieving normality acceptance rates of 98.4\%--99.9\% compared to 37\%--92\% for the generalized extreme value (GEV) distribution and 0\%--30\% for others. The spatial duration-augmented model further outperforms a nonspatial counterpart, producing spatially continuous, uncertainty-aware recovery and survival fields. Results reveal a strong south-north gradient in temporal memory: southern peninsular and island regions (Sicily, Sardinia) exhibit pronounced persistence and duration-dependent recovery hazards, whereas northern and Alpine regimes show near-zero duration dependence. Overall, drought characterization is far more sensitive to the distributional and persistence assumptions than to the RET formulation, and duration-insensitive classifications systematically  understate extended drought risk in southern Italy.

\end{abstract}


\section{Introduction}
\label{sec:introduction}
 
Drought is among the most consequential and least clearly bounded natural
hazards on Earth, developing slowly, propagating across the meteorological, agricultural,
and hydrological domains, and inducing cumulative damage on water resources,
food production, and ecosystems~\citep{van2015hydrological}. Drought severity is
not solely determined by precipitation deficits; atmospheric evaporative demand also contributes. A rise in temperatures can intensify dry conditions even where precipitation is unchanged~\citep{GRANATA2026134428}. Global assessments
of drought frequency, duration, and severity over recent decades have documented
widespread intensification of drought conditions~\citep{spinoni2014world}. Land-atmosphere feedbacks behind this behavior, especially soil moisture–temperature coupling, are
now understood to be central to how dry anomalies persist and reinforce themselves ~\citep{seneviratne2010investigating}. The Mediterranean basin has been repeatedly
identified as a climate-change hotspot in which warm trends and dry conditions are
 expected to be particularly acute~\citep{giorgi2008climate,lionello2018relation}. This
vulnerability has been made concrete by recent record heatwaves over southwestern Europe
\citep{kim20242022} and the prolonged (2021–2023) drought
that prompted a state of emergency across the Po Valley and disrupted agriculture, hydropower,
and river networks in northern Italy~\citep{pascale2025widespread}.

Italy spans a steep north–south hydroclimatic gradient, from cold, humid Alpine and Po Valley climates to warm, semiarid Mediterranean conditions in the south and on Sicily and Sardinia. Precipitation, evapotranspiration, and water availability vary sharply over short distances due to complex topography and the competing influences of Atlantic and Mediterranean circulation. This heterogeneity has motivated extensive drought research, including studies of large-scale circulation controls such as the North Atlantic Oscillation, climate-change impacts in southern Italy, cryospheric and hydrological change in the Alpine region, and the statistical properties and persistence of Italian droughts \citep{ciccarelli2008climate, ferrari2013influence, GRANATA2026134428, HERNANDEZ2025183, pascale2025widespread, peres2023dynamic, rettig2024responses}. The same 
diversity that makes Italy
scientifically valuable also makes it methodologically demanding, as models calibrated for one subregion may transfer poorly to others.
 
Quantifying drought requires standardized indices. The standardized precipitation evapotranspiration index \citep[SPEI;][]{vicente2010multiscalar} has become a primary tool because it is based on the climatic water balance between precipitation and reference evapotranspiration (RET), and is therefore sensitive to the temperature-driven evaporative demand that precipitation-only indices such as the SPI omit~\citep{seneviratne2010investigating, begueria2014standardized}. This sensitivity is crucial in hot regions, where elevated evaporative demand amplifies drought severity
\citep{begueria2014standardized}, and temperature-sensitive indices have been shown to provide more reliable drought characterizations in areas with large temperature variability~\citep{stagge2015candidate}. However, 
SPEI construction hinges on two modeling choices whose effects are often underexamined. The first is how RET is estimated: physically based formulations such as the FAO-56 Penman-Monteith equation
\citep{allen1998crop} require radiation, humidity, and wind data, whereas temperature-based methods such as Thornthwaite and Hargreaves are simpler but neglect key atmospheric controls. \citet{lee2024sensitivity} showed that both the RET method and the fitted probability distribution materially affect SPEI values, yet many studies adopt a single RET formulation without assessing the sensitivity of their conclusions to this choice.
 
The second choice concerns the probability distribution fitted to the
aggregated water-balance series before standardization. This decision is critical
because a poorly fitting distribution produces standardized values that violate
the normality the index presupposes and that misrepresent drought frequency.
In a widely cited European evaluation, \citet{stagge2015candidate} recommended
the generalized extreme value (GEV) distribution as a preferred candidate for
climatological drought indices such as SPEI. Yet the GEV is theoretically intended for block
maxima or minima~\citep{fisher1928limiting, gnedenko1943maximum} rather than for the full water-balance distribution, and the
commonly used generalized logistic (GenLog), Pearson type~III (Pe-III), and normal distributions
each impose structural constraints on skewness and tail behavior that may not
hold uniformly across a domain as varied as Italy. The importance of
distributional shape is increasingly recognized: higher-order moments,
particularly skewness, are needed to capture asymmetries in drought anomalies
and changes in the frequency and intensity of extreme, yet most
applications still focus on mean and variance alone~
\citep{GRANATA2026134428}. Classical candidate distributions also share a key
limitation: they offer only limited and often asymmetric flexibility for
representing lower and upper tails simultaneously, precisely where
drought and wet extremes occur. Recent advances address this issue. \citet{stein2021parametric} introduced a parametric bulk-and-tails 
(BATs) family that models each tail
independently via generalized Pareto-type tail indices, while retaining a well-behaved central region via a Student's t distribution. Its
suitability as a candidate distribution for SPEI construction in
heterogeneous climates has not yet been evaluated, although BATs has already shown promise in environmental applications such as nonstationary temperature modeling~\citep{KROCK2022100438}.

Index values alone cannot capture persistence, the tendency of dry conditions, once established, to continue over time. Persistence controls
drought duration, propagation, and predictability and has been studied using long-memory diagnostics such as the Hurst exponent~\citep{koutsoyiannis2003climate} and detrended fluctuation analysis (DFA)~\citep{kantelhardt2001detecting}, which quantify long-range dependence and scale-invariant
behavior in hydroclimatic series. 
 Applied to Italy~\citep{GRANATA2026134428},
these diagnostics reveal a pronounced meridional gradient with the strongest
temporal memory in the southern peninsula and islands and weaker, more rapidly fluctuating behavior in the north. Persistence has also been modeled using
Markov chains for wet and dry spells, but classical finite-order Markov chains are memoryless beyond their order and enforce geometrically distributed spell durations, contrary to the
heavy-tailed dry spells observed in Iberia and similar
regions \citep{lana2006, lopez2015}. To address this,
\citet{doize2026} proposed a duration-augmented binary Markov chain in which the probability of leaving a state depends explicitly on the
elapsed spell duration and showed its equivalence to an
alternating renewal process. This framework relaxes the memoryless assumption
without imposing a fixed memory length, but it was developed for point-scale 
rainfall occurrence and has not yet been extended to model the
spatial structure of drought persistence.

The literature review highlights three main gaps that this study addresses. First, although the sensitivity of SPEI to the RET method and to the
fitted distribution has each been examined separately
\citep{stagge2015candidate,lee2024sensitivity}, few studies assess these two
choices jointly and systematically over a heterogeneous domain. Second, classical candidate distributions used in drought
monitoring lack the flexibility to represent the asymmetric, dual-tail
behavior of the water-balance series, and the recently proposed BATs
distribution \citep{stein2021parametric} has not been evaluated as a candidate for
SPEI construction. Third, while spatial persistence in Italy has been
described through Hurst and DFA diagnostics
\citep{GRANATA2026134428}, existing duration-dependent persistence models
\citep{doize2026} treat locations independently, rely on strong parametric distribution assumptions, and have not been
embedded in a spatial framework that borrows information across space and
propagates uncertainty into persistence estimates.
 
To address these gaps, this study makes three contributions. We evaluate SPEI
constructions over the entire Italian territory using the high-resolution (0.25$^\circ \times$ 0.25$^\circ$, gridded) Meteorological variables for Agriculture:
daily time series for the Italian Area (MADIA)
agro-meteorological dataset \citep{PARISSE2023108843} and compare three RET
formulations (Penman--Monteith, Thornthwaite,
Hargreaves) to directly assess the sensitivity of drought characterization to
evaporative-demand estimation. We then introduce the
BATs distribution as a new candidate distribution for SPEI and
benchmark it against the GEV, generalized logistic, Pearson type~III, and
normal distributions using normality-based goodness-of-fit testing at
accumulation scales from one to twelve months. Finally, we develop a Bayesian
spatial extension of the duration-augmented Markov chain framework of~\citet{doize2026}. Here, Gaussian process priors allow the duration
dependence of drought recovery to vary smoothly across space, yielding
spatially resolved, uncertainty-aware estimates of drought persistence that
complement and formalize descriptive persistence gradients. Through these contributions, the study provides a flexible
and a theoretically coherent basis for drought monitoring and deepens our 
understanding of extreme droughts across Italy.

The remainder of this paper is organized as follows. Section~\ref{sec:methods}
describes the study area and the dataset, the three
RET formulations, the candidate distributions
considered for SPEI construction, including the BATs as new candidate distribution, the
goodness-of-fit evaluation measures, and the Bayesian spatial
duration-augmented Markov chain framework to model drought persistence.
Section~\ref{sec:results} presents the results, covering the spatial
distribution of RET, the comparative performance of
the candidate distributions across accumulation scales, the resulting drought
trends and long-range-dependence diagnostics over Italy, and the spatially
resolved persistence and recovery and survival estimates obtained from the fitted spatial
model. Finally, Section~\ref{sec:conclusion} concludes the paper and highlights future working directions.

\section{Study area and methodology}\label{sec:methods}
\subsection{Study area and dataset}

This study considers the entire Italian territory, which is distinguished by its considerable diversity in climate, topography, and agro-environmental conditions. The territory stretches from the Alpine arc in the north to the Mediterranean basin in the south, covering approximately 301,000 km$^2$. Such geographic heterogeneity gives rise to well-defined climatic gradients, spanning the cold, humid alpine and subcontinental regimes of the north; the temperate and semiarid Mediterranean conditions of the central and southern regions; and the coastal environments marked by mild winters and hot, arid summers~\citep{dibari2021climate}.
The heterogeneity between climate and terrain directly influences Italy's agricultural systems and hydrological dynamics. 
Northern Italy, including the Po Valley constitutes one of Europe's most productive agricultural regions, sustained by extensive irrigation networks and fertile alluvial soils~\citep{soana2024climate}. 
Central Italy presents a mosaic of plains and rolling hills 
where cereals, olive trees, and vineyards are cultivated on a large scale, while southern Italy and the major islands (Sicily and Sardinia) experience warmer and drier conditions that promote higher evapotranspiration and greater vulnerability to drought~\citep{chiappini2024olive, granata2025hydrological}. 

\begin{figure}[h]
\includegraphics[width=1\textwidth,height=10cm]{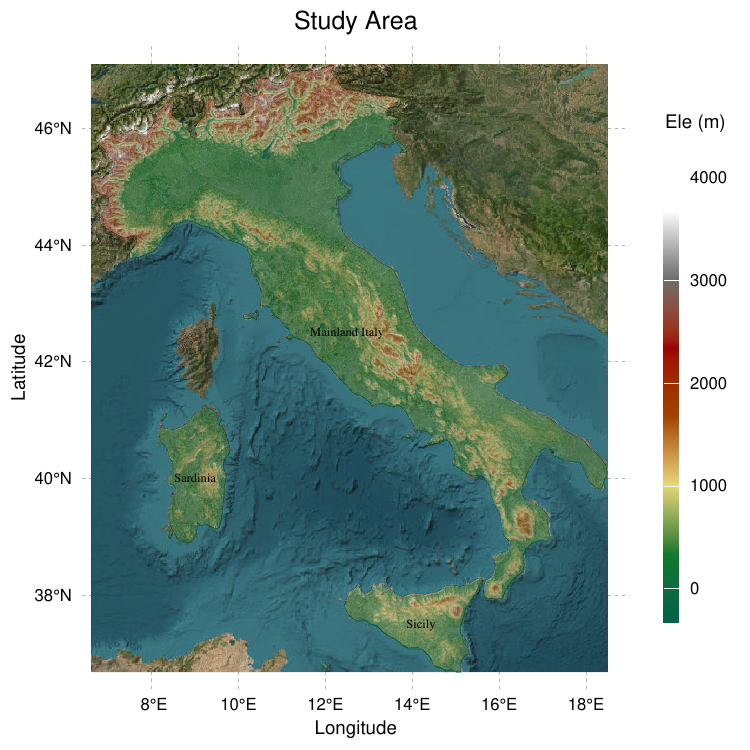}
    \caption{Study area spatial map.}
    \label{fig:st_map}
\end{figure}

The analysis employs the MADIA dataset \citep{PARISSE2023108843}, which provides a consistent, high-resolution database of agro-meteorological variables across Italy, derived from hourly ERA5 surface reanalysis data. 
The dataset covers the period 1981--2022 at a spatial resolution of 0.25$^\circ \times$ 0.25$^\circ$ and provides daily time series of the variables listed in Table~\ref{data-variables} and a spatial map of the area in Figure~\ref {fig:st_map}. 
\begin{table}[h!]
\centering
\caption{Meteorological variables used in the study}
\begin{tabular}{ll}
\hline
\textbf{Variable} & \textbf{Description / Unit} \\
\hline
$T_\mathrm{min}$ & Minimum air temperature ($^\circ$C) \\
$T_\mathrm{mean}$ & Mean air temperature ($^\circ$C) \\
$T_\mathrm{max}$ & Maximum air temperature ($^\circ$C) \\
RH$_\mathrm{min}$ & Minimum relative humidity (\%) \\
RH$_\mathrm{max}$ & Maximum relative humidity (\%) \\
WS10 & Wind speed at 10 m above ground level (m s$^{-1}$) \\
$R_\mathrm{s}$ & Downward shortwave solar radiation (MJ m$^{-2}$ day$^{-1}$) \\
P & Precipitation (mm day$^{-1}$) \\
$\text{RET}$ & Reference evapotranspiration (mm day$^{-1}$), computed using \\ &FAO Penman–Monteith~\citep{allen1998crop} \\
\hline
\end{tabular}\label{data-variables}
\end{table}

The broad spatial and temporal coverage of MADIA enables a comprehensive analysis of the climatic variability that governs agricultural water balance and drought conditions across Italy. The dataset captures both the strong north--south gradient in temperature, humidity, and precipitation and the distinct seasonal cycles typical of Mediterranean climates. Temperature, solar radiation, and reference evapotranspiration reach their maxima during the summer months, whereas relative humidity, wind speed, and precipitation tend to peak in winter.

\begin{figure}
    \centering
    \subfigure{
\includegraphics[width=1\textwidth,height=8cm]{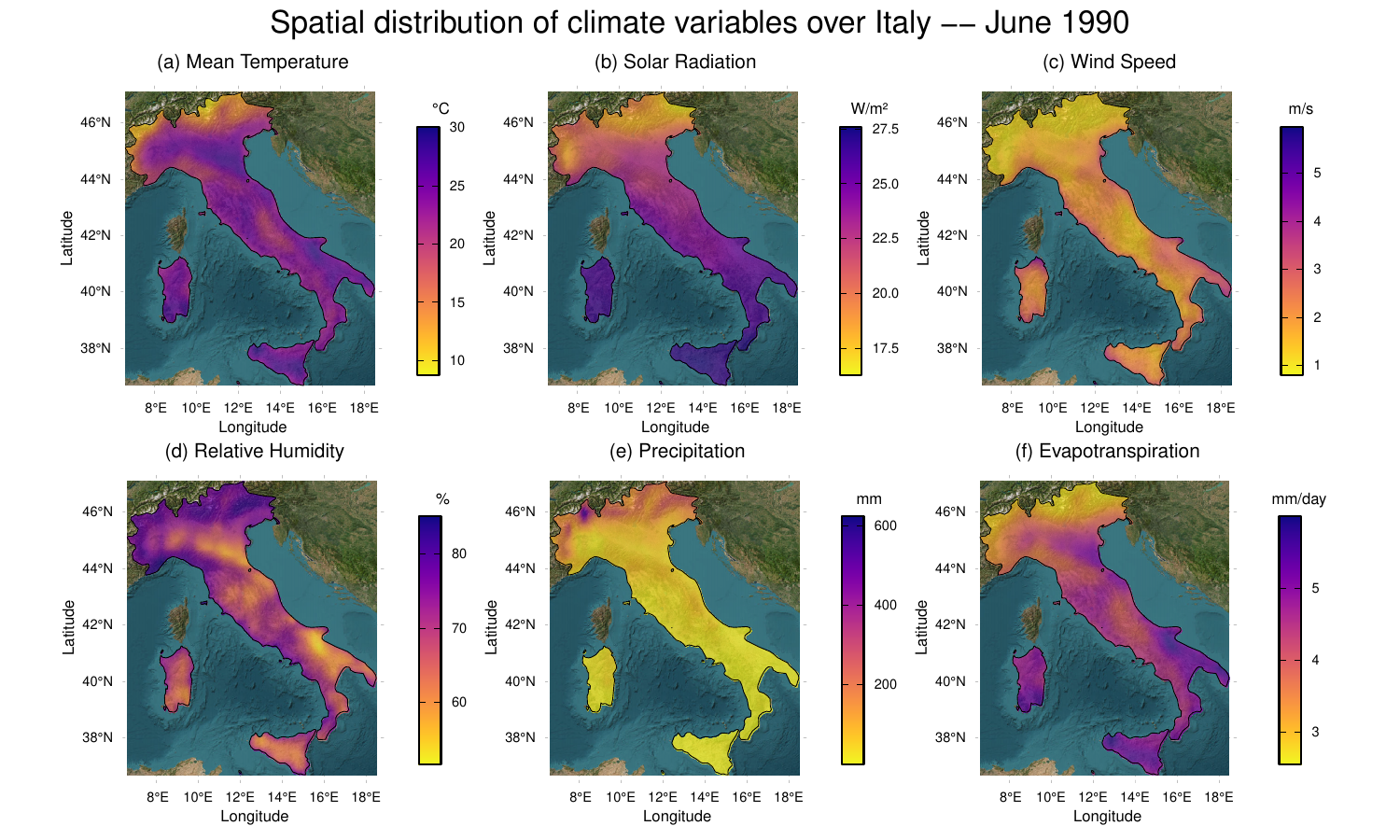}
    }
    \subfigure{
\includegraphics[width=1\textwidth,height=8cm]{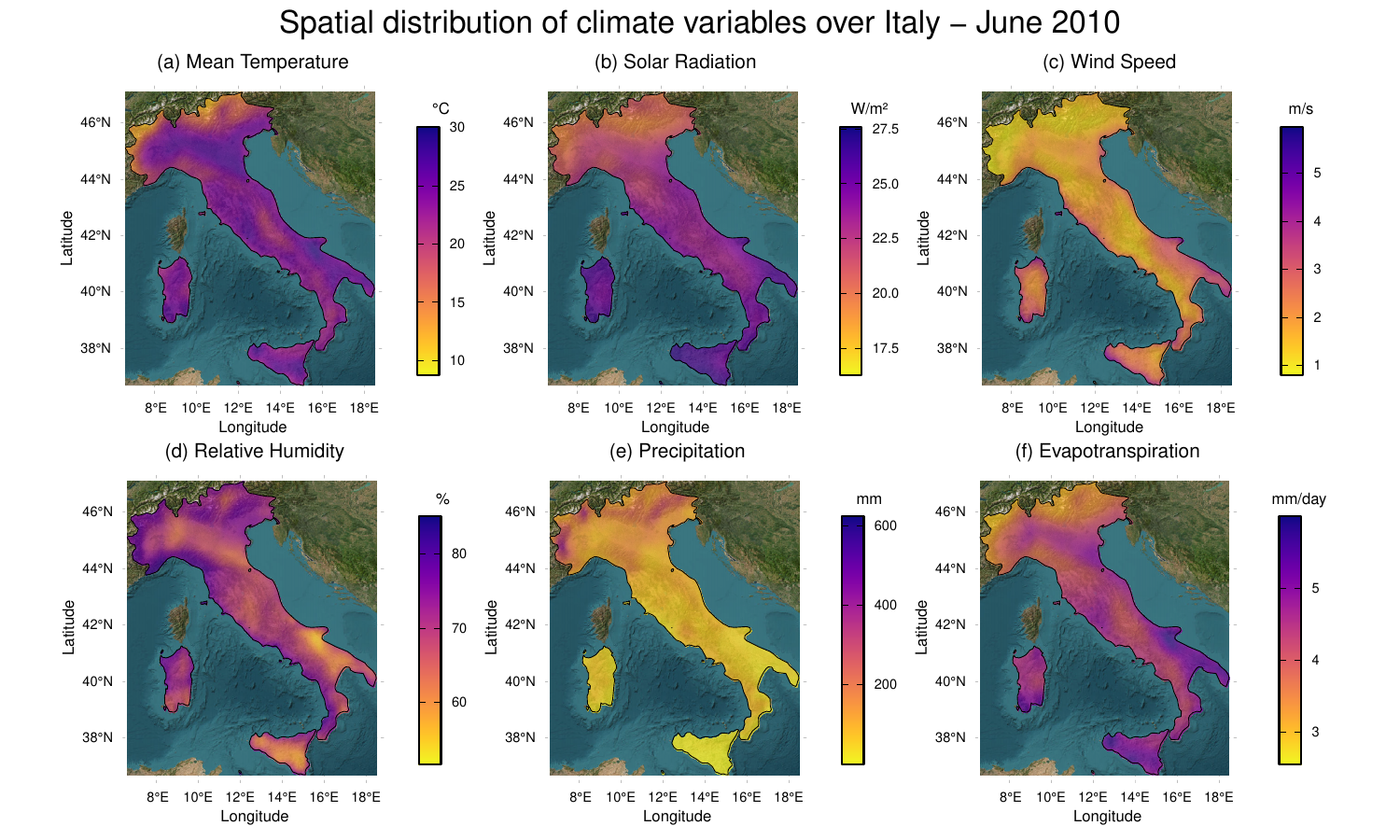}
    }
    \caption{Spatial distribution of variables for June 1990 and June 2010: (a) mean maximum temperature, (b) mean solar radiation, (c) mean wind speed, (d) mean relative humidity, (e) precipitation sum, and (f) mean FAO Penman–Monteith RET. }
    \label{fig:spatial_dist}
\end{figure}

The spatial distributions of the SPEI input variables in June 1990 and June 2010~(Figure~\ref{fig:spatial_dist})
for June 1990 and June 2010 
reveal substantial hydroclimatic variability across the region. The variables examined include mean air temperature, solar radiation, wind speed, relative humidity, precipitation, and RET estimated using the FAO56 Penman–Monteith method. All exhibit strong spatial variability, reflecting Italy's complex topography and Mediterranean climatic conditions.

A clear north-to-south variability appears in both years. Northern Italy, particularly the Alpine and Po Valley regions, shows lower temperatures, higher precipitation, and higher relative humidity, whereas southern Italy and the islands are warmer and drier, with higher solar radiation and evapotranspiration. This reflects the Mediterranean summer climate, in which the south experiences stronger evaporative demand and reduced moisture availability~\citep{vavassori2026recent}.
Temperature increases from north to south. Mountainous regions remain cooler due to elevation effects, while the southern peninsula and coastal zones exceed 25--30$^\circ$C in several areas. Solar radiation follows the same pattern, with the south receiving more energy from reduced cloud cover and stronger insolation, further enhancing evapotranspiration. Relative humidity varies inversely with both temperature and moisture content; the cooler, wetter north remains more humid, while the south is dominated by drier air masses. Wind speed is comparatively uniform, with only slightly higher values in coastal and southern areas linked to Mediterranean circulation and sea breezes~\citep{laurila2021climatology}.
Precipitation shows the strongest spatial contrast. The north receives considerably higher rainfall, mainly from Alpine orographic effects and enhanced moisture transport, whereas the south records markedly lower totals consistent with Mediterranean summer dryness~\citep{moccia2021spatial}. RET closely follows temperature and solar radiation, with higher values in southern and central Italy, where high temperatures, strong radiation, low humidity, and moderate winds raise evaporative demand, and lower values in the cooler, more humid north.
The two years share a broadly consistent spatial structure and the same dominant north-south variability. The main difference is that 2010 appears slightly warmer and drier across several southern and central regions, with modest increases in evapotranspiration—likely reflecting interannual variability and possible warming.
This variability is central to evapotranspiration drought indices, which are based on the climatic water balance between precipitation and RET~\citep{hui2018drought}. Regions combining low precipitation with high evapotranspiration, most notably southern Italy, are therefore the most drought-prone, whereas the northern regions remain more favorable, receiving higher rainfall and lower evaporative demand.

\subsection{Reference evapotranspiration equations}\label{ret-eq}

To derive the daily $\mathrm{RET}$ values, we used three different equations. First, the  $\mathrm{RET}$ was calculated using the modified FAO–56 Penman–Monteith equation~\citep{allen1998crop}. The modified Penman–Monteith  $(\text{RET})$ equation is
\begin{equation}\label{PM}
\mathrm{RET_{PM}} = \frac{0.408\,\Delta (\mathcal{R}_n - \mathcal{G}) + \eta \frac{900}{\mathcal{T} + 273} \, \delta_2 (\omega_s - \omega_a)}{\Delta + \eta (1 + 0.34 \delta_2)}
\end{equation}

\noindent
where $\mathrm{RET_{PM}}$ is the refenece evapotranspiration (mm day$^{-1}$), 
$\mathcal{R}_n$ is the net radiation (MJ m$^{-2}$ day$^{-1}$), 
$\mathcal{G}$ is the soil heat flux density (MJ m$^{-2}$ day$^{-1}$), 
$\eta$ is the psychrometric constant (kPa $^\circ$C$^{-1}$), 
$\mathcal{T}$ is the mean daily air temperature ($^\circ$C), 
$\delta_2$ is the wind speed at 2 m height (m s$^{-1}$), 
$\Delta$ is the slope of the saturation vapour pressure curve (kPa $^\circ$C$^{-1}$), 
and $(\omega_s - \omega_a)$ is the saturation vapour pressure deficit (kPa), for more details, see e.g., \citet[Appendix A]{stagge2015candidate}.

Daily values were aggregated into monthly values to compute the monthly climatic water balance. Based on monthly aggregation, the Thornthwaite-based RET equation is defined as 
\begin{equation}\label{thrwait}
\mathrm{RET_{TW}} = 16\rho\left(\frac{10\mathcal{T}}{\mathcal{I}}\right)
\end{equation}
where $\mathcal{I}$ is the heat index for the whole year, and $m$ is a coefficient as a function of $\mathcal{I}$ such that $
m = 6.75 \times 10^{-7} \mathcal{I}^3 - 7.71 \times 10^{-5} \mathcal{I}^2 + 1.79 \times 10^{-2} \mathcal{I} + 0.492
$
and $\rho$ is a correction coefficient as a function of latitude and month~\citep{lee2024sensitivity}. The Hargreaves-based $\mathrm{RET}$ equation is 
\begin{equation}\label{HG}
\mathrm{RET_{HG}} = 0.0023\,\mathcal{R}_a (\mathcal{T} + 17.8) (\mathcal{T}_{\max} - \mathcal{T}_{\min})^{0.5}
\end{equation}

\noindent
where $\mathcal{T}_{\max}$ and $\mathcal{T}_{\min}$ are the maximum and minimum air temperatures ($^\circ$C), respectively, and $\mathcal{R}_a$ is the extraterrestrial radiation (MJ m$^{-2}$).

The RET values from the above three equations against different time scales are calculated later in the results Section~\ref{sec:results} by supplying the necessary MADIA data variables.

\subsection{SPEI and candidate distributions}
To calculate the SPEI, the monthly climatic water balance deficit $\mathcal{D}$ will first be computed at each grid point
as the difference between precipitation and $\mathrm{RET}_{j}, j\in \{\mathrm{PM, TW, HG}\}$ as 
\begin{equation}\label{d-constrction}
    {\mathcal{D}_{j}}= P-\mathrm{RET}_{j}
\end{equation}
In order to evaluate the 
sensitivity of SPEI to the choice of $\mathrm{RET}$, all three methods described in~Section~\ref{ret-eq} are considered in \eqref{d-constrction}.


Once the monthly deficit series $D_j$ is obtained, the data are aggregated over 
different time scales (e.g., $1, 2, \dots,  $months) to capture short-term and 
long-term moisture conditions. These aggregated series are then used to represent 
the accumulated water balance at each grid point over the Italian territory.

A parametric probabilistic model is subsequently required to be fitted to the aggregated $\mathcal{D}_j$ time series at each grid point. To model the distribution $\mathcal{D}_j$ with flexible tail behavior and a well-characterized central region, we adopted the novel BATs candidate 
 distribution~\citep{stein2021parametric}. The acronym “BATs” here explicitly reflects the modeling of dual-tail formulation. The BATs model introduces a parametric family of distributions that captures both heavy and bounded tails, offering improvements over classical bulk-and-tail models, which often lack flexibility in the lower tail. The practical advantages of BATs allow us to flexibly accommodate both drought and wet extremes without sacrificing a stable central behavior. To make it easier for the reader, we move the technical details regarding the BATs model construction to Appendix~\ref{add_Bats}. We developed an \texttt{R} script to estimate the BATs model using the maximum likelihood approach, which is available on \texttt{GitHub}\footnote{https://github.com/touqeerahmadunipd/SDAP}.

\noindent\textbf{Existing candidate distributions:} In comparison to the BATs model, four widely used probability distributions will be considered subsequently for $\mathcal{D}_j$ in order to construct SPEI: the GEV, GenLog, Pe-III, and Normal distributions.
These distributions have been widely used in drought characterization and hydroclimatic studies due to their ability to capture the statistical properties of climatic water balance data. The GenLog and Pe-III distributions are commonly employed in drought frequency analysis due to their flexibility in capturing the skewness of hydrological and climatic variables. The Normal distribution, although computationally simple and widely used, may not adequately represent the asymmetry and heavy-tail characteristics often observed in $\mathcal{D}_j$ series.
The GEV distribution has also been investigated in several drought and climate studies due to its strong ability to model skewed and extreme behavior in hydroclimatic datasets~\citep{stagge2015candidate, lee2024sensitivity}. However, applying the GEV to $\mathcal{D}_j$ raises an important conceptual and theoretical consideration. The GEV distribution is theoretically developed to model block maxima or minima~\citep{fisher1928limiting, gnedenko1943maximum}, whereas $\mathcal{D}_j$ represents the full distribution of climatic water balance conditions, not just extreme events. 
Nevertheless, previous studies have demonstrated that GEV can still provide satisfactory empirical performance in fitting water-balanced data, particularly in regions (e.g., all over Europe) where drought series exhibit pronounced skewness and heavy-tail behavior~\cite{stagge2015candidate}.
Accordingly, these four distributions were selected as benchmark models to comprehensively evaluate the performance, robustness, and suitability of the BATs model for SPEI calculation and drought assessment. 

\subsection{Evaluation measures}
The calculated SPEI$_{j}, j\in \{\mathrm{PM, TW, HG}\}$ from different candidate distributions for different temporal scales is expected to follow a standard normal distribution with $\mu=0$ and $\sigma=1$. The evaluation is performed on the final standardized SPEI values rather than directly on the fitted probability distributions. The Shapiro–Wilk (SW) test~\citep{Shapiro1965} is applied to assess the normality of the computed SPEI values because of its strong statistical power for detecting departures from normality, particularly for small and moderate sample sizes. 
This test is not strictly specific to extreme tails; while it globally provides a distribution-free assessment of whether the derived SPEI series satisfies the normality assumption required for robust drought characterization. Before applying the tests, the SPEI series was checked for temporal autocorrelation to verify that the assumptions of independence and identical distribution underlying the tests were reasonably satisfied. The null hypothesis assumes that the SPEI values are normally distributed. 
Because normality was assessed across multiple SPEI series, the resulting $p$-values were adjusted using the Benjamini-Hochberg false discovery rate procedure~\citep{benjamini1995controlling}. Let $p_{(1)} \leq p_{(2)} \leq \cdots \leq p_{(m)}$ denote the ordered $p$-values from the $m$ normality tests. The corresponding adjusted $p$-values are computed as
\[
\tilde{p}_{(i)}=\min\left\{\min_{j\geq i}\left(\frac{m}{j}p_{(j)}\right),\,1\right\},
\]
where $m$ is the total number of tests. A significance level of $\alpha=0.05$ was adopted. Therefore, adjusted $p$-values greater than $0.05$ indicate that the null hypothesis cannot be rejected, suggesting that the corresponding SPEI series adequately satisfies the normality assumption. Conversely, adjusted $p$-values less than or equal to $0.05$ indicate statistically significant departures from normality.



The temporal trends of SPEI$_{j}$, where $j \in {\mathrm{PM, TW, HG}}$, were evaluated using the Mann–Kendall trend test. Similar to the SW test, the SPEI time series at each grid cell was examined for temporal autocorrelation. If significant autocorrelation was present, the modified Mann–Kendall test proposed by \citet{hamed1998} was applied by adjusting the original variance as
\begin{equation}
\Var_{m}(S) = \Var_{o}(S)\Bigg[1 + \frac{2}{n(n-1)(n-2)}
\sum_{i=1}^{n-1} (n-i)(n-i-1)(n-i-2) r_i \Bigg],
\end{equation}
where $\Var_{m}(S)$ and $\Var_{o}(S)$ denote the modified and original variances of the Mann–Kendall statistic $S$, respectively, $n$ is the sample size, and $r_i$ represents the autocorrelation coefficient at lag $i$~\citep{ng2024assessment}. A monotonic trend was considered statistically significant at the 5\% significance level when the absolute value of the standardized test statistic satisfied $|Z| > 1.96$.
In addition, the nonparametric Sen’s slope estimator was applied to quantify the magnitude and direction of the detected trends. The rate of change of the trend, denoted as $R_{\mathrm{Sen}}$, was calculated as
\begin{equation}
R_{\mathrm{Sen}} =
\mathrm{median}
\left(
\frac{x_j - x_i}{j-i}
\right),
\qquad
1 \leq i < j \leq n,
\end{equation}
where $x_i$ and $x_j$ represent the values of the variable at time steps $i$ and $j$, respectively, and $n$ is the total number of observations. Positive values of $R_{\mathrm{Sen}}$ indicate an increasing trend, whereas negative values indicate a decreasing trend.

\subsection{Bayesian spatial drought persistence modelling}\label{Bayesian_model}

Our drought persistence modelling approach is motivated by the Binary
Markov Chain with Duration (BMCD) framework introduced
by~\citet{doize2026} for rainfall occurrence.  The key insight of the
BMCD is that the transition probability out of a given state
(e.g.\ drought in our case) depends explicitly on the current spell
duration, thereby relaxing the memoryless property inherent to
finite-order Markov chains~\cite{stroock2005introduction}.  \citet{doize2026} establish a formal
equivalence between the BMCD representation and an alternating renewal
process, demonstrating that spell durations become independent and
identically distributed under this construction.  This equivalence
provides a principled foundation for modelling duration-dependent
persistence without imposing an arbitrary upper bound on the order of
memory, a limitation that affects standard hybrid-order Markov chains.

Let \(X_t(s) \in \{0, 1\}\) denote the drought state at location
\(s \in \mathcal{S}\) and time~\(t\), where \(X_t(s) = 1\) indicates
drought conditions
(SPEI\(_j < -1\), where \(j \in \{\text{PM}, \text{TW}, \text{HG}\}\)
denotes the PM, TW, and HG reference evapotranspiration methods, respectively, each evaluated at
accumulation timescales of \(1, 2, \ldots\) months) and
\(X_t(s) = 0\) indicates non-drought conditions.  Define \(D_t(s)\)
as the elapsed duration of the ongoing spell at time~\(t\), i.e.\ the
number of consecutive months the process has remained in state
\(X_t(s)\).  Following \citet{doize2026}, the pair
\(\bigl(X_t(s),\, D_t(s)\bigr)\) forms a Markov chain with state
space \(\{0,1\} \times \mathbb{N}^*\).  The transition dynamics are
governed by the exit probabilities
\[
  q_d^{(r)}(s)
    = \mathbb{P}\!\bigl(X_{t+1}(s) = 1 - r \mid X_t(s) = r,\;
                        D_t(s) = d\bigr),
  \qquad r \in \{0, 1\},
\]
which represent the probability of leaving state~\(r\) after having
occupied it for \(d\) consecutive time steps.  In the classical
two-state first-order Markov chain, \(q_d^{(r)}\) is constant with
respect to~\(d\), forcing geometrically distributed spell durations.
The BMCD relaxes this restriction, allowing \(q_d^{(r)}\) to vary
with~\(d\) and thus accommodating a much richer class of spell
duration distributions, including those with heavy tails that
characterise persistent dry spells~\citep{lana2006, lopez2015}.

For a given location~\(s\), let \(\tau^{(0)}(s)\) denote the random
duration of a drought spell (state \(r = 0\) in the notation of
\citet{doize2026}, corresponding to \(X = 1\) in our drought
indicator).  The exit probability at duration~\(d\) is related to the
survival function of \(\tau^{(0)}(s)\) by~\citep[Eq.~(4)]{doize2026}
\begin{equation}
  q_d^{(0)}(s)
    = \frac{\mathbb{P}\!\bigl(\tau^{(0)}(s) = d\bigr)}
           {\mathbb{P}\!\bigl(\tau^{(0)}(s) \ge d\bigr)}
    = \frac{\overline{F}_{\tau^{(0)}(s)}(d-1)
            - \overline{F}_{\tau^{(0)}(s)}(d)}
           {\overline{F}_{\tau^{(0)}(s)}(d-1)}
    = 1 - \frac{\overline{F}_{\tau^{(0)}(s)}(d)}
               {\overline{F}_{\tau^{(0)}(s)}(d-1)},
\end{equation}
where \(\overline{F}_{\tau^{(0)}(s)}(d) =
\mathbb{P}\!\bigl(\tau^{(0)}(s) > d\bigr)\) is the survival function.
This relationship is fundamental; specifying a parametric distribution
for spell durations directly determines the duration-dependent hazard
of drought termination.  Conversely, modeling the hazard function
implies a particular spell duration distribution.  This duality,
while implicit in the alternating renewal
representation~\citep{resnick1992}, has been systematically exploited
in the drought modelling context only recently.

We adopt the latter approach, modelling the discrete-time recovery
hazard directly as a function of duration and spatial covariates.
Define the recovery hazard as
\[
  h_t(s)
    = \mathbb{P}\!\bigl(X_{t+1}(s) = 0 \mid X_t(s) = 1,\; D_t(s) = d\bigr)
    = q_d^{(0)}(s),
\]
where \(r = 0\) in the notation of \citet{doize2026} denotes the
drought (dry) state.  To ensure that observations are conditional on
being in an active drought spell, we restrict the dataset to time
steps for which \(X_{t-1}(s) = 1\).  This conditioning is essential:
the hazard is defined only while the drought persists, and including
non-drought periods would confound the estimation of duration
dependence~\citep{singer2003}.

Furthermore, to avoid boundary-censoring artefacts induced by the
arbitrary start and end of the observational record, the first and
last drought spells within each continuous data segment are excluded
from the analysis, as also adopted follow by~\citep{doize2026}.  This ensures that
only spells with a clearly observed onset and a clearly observed
termination contribute to the likelihood.

\subsubsection{Spatial hazard specification}

We model the logit-transformed recovery hazard as an additive function
of duration and spatial location:
\begin{equation}
  \operatorname{logit}\!\bigl(h_t(s)\bigr)
    = \alpha_0 + \alpha(s)
      + \bigl[\beta_0 + \beta(s)\bigr] \cdot \log\!\bigl(D_t(s) + 1\bigr),
\end{equation}
where \(\alpha_0\) is a global intercept (fixed effect) representing
the baseline log-odds of recovery at duration \(D = 1\) after
accounting for spatial variation; \(\beta_0\) is a global slope
(fixed effect) capturing the average effect of log-duration on
recovery odds; \(\alpha(s)\) is a zero-mean Gaussian process (GP)
capturing spatial deviations in the baseline hazard; and \(\beta(s)\)
is a zero-mean GP capturing spatial deviations in duration-dependence.

This formulation separates global (domain-average) behavior from
local spatial adjustments.  The total slope
\(\beta_{\text{total}}(s) = \beta_0 + \beta(s)\) determines whether a
location exhibits persistent drought behavior (negative total slope,
recovery hazard declining with duration) or self-terminating behavior
(positive total slope).  In our spatial extension of the BMCD
framework, this decomposition allows the persistence regime to vary
continuously across the study domain while pooling information across
nearby locations through the GP prior.
\\
\medskip
\textbf{Gaussian Process Priors:}
Spatial dependence is modeled using independent zero-mean Gaussian
process priors for \(\alpha(s)\) and \(\beta(s)\).  For a set of
locations \(s_1, \dots, s_n\),
\[
  \boldsymbol{\alpha} \sim \mathcal{N}\!\bigl(\mathbf{0},\,
    \sigma_\alpha^2\,\mathbf{R}_\alpha\bigr),
  \qquad
  \boldsymbol{\beta} \sim \mathcal{N}\!\bigl(\mathbf{0},\,
    \sigma_\beta^2\,\mathbf{R}_\beta\bigr),
\]
where the correlation matrices \(\mathbf{R}\) have entries given by
the Mat\'{e}rn~5/2 covariance function
\[
  R_{ij}
    = \left(1 + \frac{\sqrt{5}\,d_{ij}}{\rho}
               + \frac{5\,d_{ij}^2}{3\rho^2}\right)
      \exp\!\left(-\frac{\sqrt{5}\,d_{ij}}{\rho}\right),
\]
with \(d_{ij}\) the Euclidean distance between locations \(s_i\) and
\(s_j\) in degree space and \(\rho\) the range parameter controlling
the decay of spatial correlation.  The Mat\'{e}rn~5/2 kernel is
chosen because it produces sample paths that are twice differentiable,
a smoothness level appropriate for smoothly varying environmental
fields~\citep{stein1999}.  Separate range parameters \(\rho_\alpha\)
and \(\rho_\beta\) allow the spatial scales of the baseline hazard and
duration-dependence to differ.

Large-scale GP inference is made computationally tractable via the
Hilbert Space GP (HSGP) approximation~\citep{riutort2023practical}, which
represents each GP using \(m\)
basis functions derived from the
eigenfunctions of the Laplacian on a bounded domain, reducing
computational complexity from \(\mathcal{O}(n^3)\) to
\(\mathcal{O}(nm)\).

\noindent\textbf{Prior Specification:}
Prior distributions are chosen to be weakly informative but
constrained by physical reasoning and computational considerations.\\
\noindent\textit{Fixed effects:}
Following the recommendations for logistic regression~\citep{gelman2008},
we set \(\alpha_0 \sim \mathcal{N}(0,\, 2^2)\) 
and \(\beta_0 \sim \mathcal{N}(0,\, 1^2)\).
These priors assign high probability to
plausible odds-ratio scales while avoiding the extreme predictions that
arise from improper flat priors.\\
\noindent\textit{GP marginal standard deviations:}\quad
For both \(\sigma_\alpha\) and \(\sigma_\beta\) (the GP marginal
\emph{standard deviations}), we use an inverse-gamma prior
\(\operatorname{Inv\text{-}Gamma}(2.5,\, 0.5)\).  The mean of this
prior is \(0.5/(2.5 - 1) \approx 0.33\) on the standard-deviation
scale, placing most mass in the range 0.1--1 on the logit-transformed
hazard scale.  This allows the data to dominate while keeping spatial
variation plausible.  The inverse-gamma is a common choice for
variance components in hierarchical
models~\citep{gelman2006}.
\\
\noindent\textit{GP length scales.}\quad
The range parameters \(\rho_\alpha\) and \(\rho_\beta\) receive
independent \(\operatorname{Gamma}(2,\, \lambda)\) priors, with the
rate~\(\lambda\) calibrated to the spatial extent of the study domain.
Specifically, if \(L\) denotes the average of the longitudinal and
latitudinal ranges of the station network, we set the prior mean to
\(0.25 \times L\) (approximately 25\% of the domain extent).  The
\(\operatorname{Gamma}(2,\, \lambda)\) prior has a shape parameter of
2, which places more mass near the mean than an exponential prior,
reflecting a modest prior belief that spatial correlation is
meaningful.

\subsubsection{Bayesian Inference}

The model is estimated using Hamiltonian Monte Carlo (HMC) as
implemented in Stan~\citep{carpenter2017} via the \texttt{brms}
interface~\citep{buerkner2017}.  Four chains are run for 2{,}000
iterations each, discarding the first 1{,}000 as warm-up, yielding
4{,}000 post-warm-up draws in total.  Convergence is assessed using
the \(\hat{R}\) statistic (target \({<}\,1.01\)) and bulk effective
sample size (ESS \({>}\,400\) per parameter).  Posterior predictive
checks compare observed recovery rates to replicate data simulated
from the fitted model, with discrepancies summarised by a Bayesian
\(p\)-value~\citep{gelman2013}.

\subsubsection{Reconstruction of Spatial Persistence Fields}

From the posterior draws, we recover the total slope field
\(\beta_{\text{total}}(s) = \beta_0 + \beta(s)\) directly from the
linear predictor structure.  For each location~\(s\), evaluate the
linear predictor at two reference durations:
\[
  \eta(s, d)
    = \alpha_0 + \alpha(s)
      + \bigl(\beta_0 + \beta(s)\bigr) \cdot \log(d + 1).
\]
The difference \(\eta(s, 1) - \eta(s, 0)
= \bigl(\beta_0 + \beta(s)\bigr) \cdot \log 2\), so $
  \beta_{\text{total}}(s)
    = \frac{\eta(s, 1) - \eta(s, 0)}{\log 2}.
$
This computation is performed for each posterior draw, yielding a full
posterior distribution for \(\beta_{\text{total}}(s)\).  The
proportion of draws with \(\beta_{\text{total}}(s) < 0\) provides a
posterior probability that the location~\(s\) exhibits persistent drought
behavior (recovery hazard declines with duration).  This approach
propagates all sources of uncertainty and estimation uncertainty in
fixed effects and full spatial uncertainty in the GP fields into
the final persistence classification, avoiding the overconfidence that
would result from point-estimate plug-in
methods~\citep{gelman2006}.



\noindent \textbf{Recovery and Survival Probability Curves:} For each location~\(s\) and duration \(d = 1, \dots,\)  months, the
posterior recovery probability for draw~\(k\) is
\[
  h^{(k)}(s, d)
    = \operatorname{logit}^{-1}\!\bigl(\eta^{(k)}(s, d)\bigr),
    \qquad k = 1, \dots, K,
\]
where \(\eta^{(k)}(s, d)\) is the draw-specific linear predictor
evaluated at duration~\(d\). The posterior mean
\(\hat{h}(s, d)\) and pointwise 95\% credible intervals are obtained
by summarizing across draws. These recovery probabilities represent
the conditional probability that a drought event terminates at
duration~\(d\), given that it has persisted up to duration~\(d-1\).
Posterior survival probabilities are derived directly from the fitted
Bayesian hazard model as,
\[
  S^{(k)}(s,d)
  =
  \prod_{j=1}^{d}
  \bigl(1-h^{(k)}(s,j)\bigr),
\]
which represents the probability that a drought event persists longer
than duration~\(d\). 
Slowly declining survival
curves indicate persistent drought regimes, whereas rapidly declining
curves correspond to short-lived drought events. 

\section{Results}\label{sec:results}

\subsection{Reference evapotranspiration}

\begin{figure}[t]
    \centering
\includegraphics[width=1\textwidth,height=8cm]{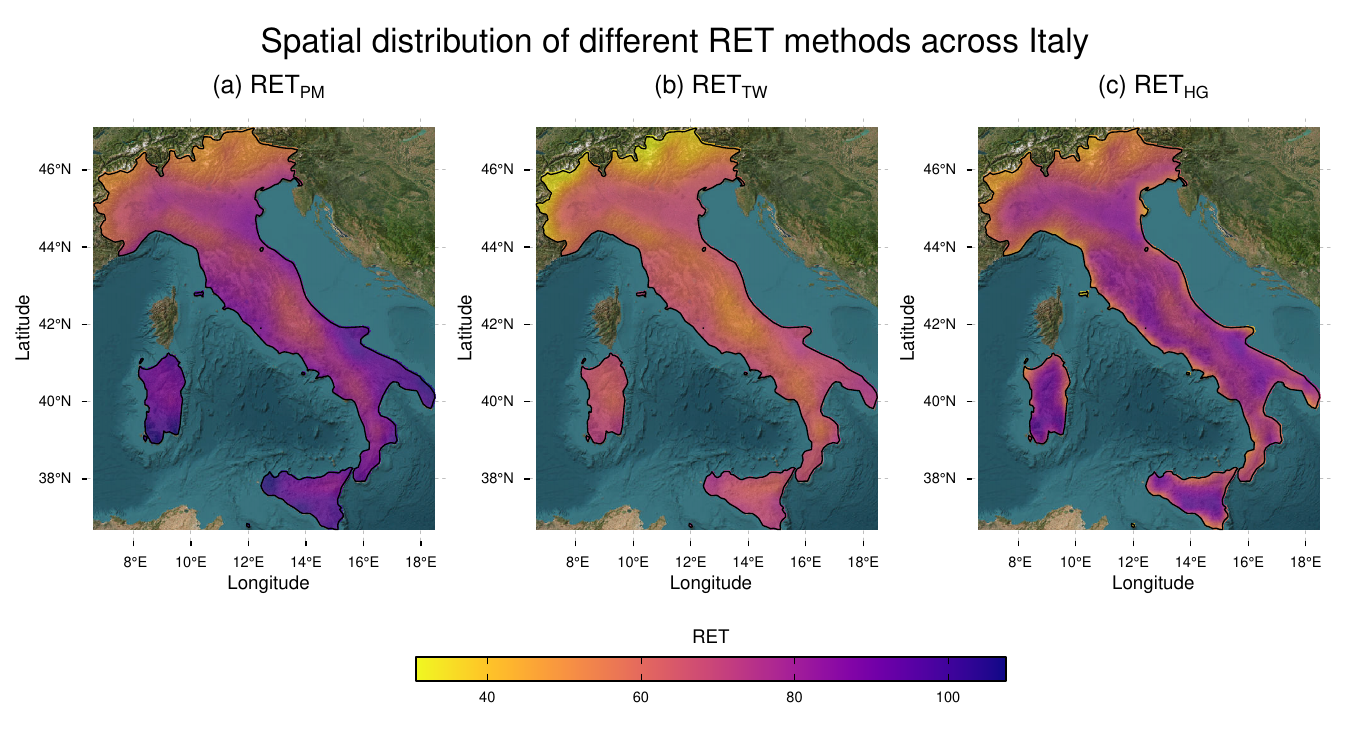}
    \caption{Spatial distribution of  $\mathrm{RET}_{j}$, $j \in {\mathrm{PM, TW, HG}}$ at one  month scale.}
    \label{fig:ret_spatial}
\end{figure}

The monthly $\mathrm{RET}_{j}$, $j \in {\mathrm{PM, TW, HG}}$, was calculated using Eqs.~\eqref{PM}, \eqref{thrwait}, and \eqref{HG} at each of 730 grid cells across Italy. Long-term mean monthly RET values were computed at each grid cell to investigate their spatial distribution (Figure~\ref{fig:ret_spatial}, a--c). The three RET methods revealed a broadly consistent spatial pattern with lower RET values over northern Italy and mountainous regions and progressively higher RET values toward southern Italy and the Mediterranean coastal areas. This spatial heterogeneity primarily reflects the influence of temperature, solar radiation, and elevation on atmospheric evaporative demand. 

Although the general spatial structure was similar among the three methods, important differences in RET magnitude and spatial variability were observed. The $\mathrm{RET_{PM}}$ and $\mathrm{RET_{HG}}$ approaches exhibited greater spatial variability and yielded higher RET estimates across most of Italy, with the strongest effects in southern regions and along coastal zones. In contrast, $\mathrm{RET_{TW}}$ produced lower RET estimates and relatively smoother spatial transitions, especially across the Alpine and semi-arid regions. 
Figure~\textcolor{red}{S.1} shows spatial distribution of the differences of $\mathrm{RET_{TW}}$ and $\mathrm{RET_{HG}}$ with $\mathrm{RET_{PM}}$. $\mathrm{RET_{TW}}$ is seen to be underestimated relative to the $\mathrm{RET_{PM}}$ method because its temperature-based formulation does not explicitly account for other atmospheric controls, such as humidity, radiation, and wind speed~\textcolor{red}{\citep{stagge2015candidate}}, while $\mathrm{RET_{HG}}$ showing slightly smaller differences in coastal areas and the Puglia region. 

\begin{figure}
    \centering
\includegraphics[width=1\textwidth,height=8cm]{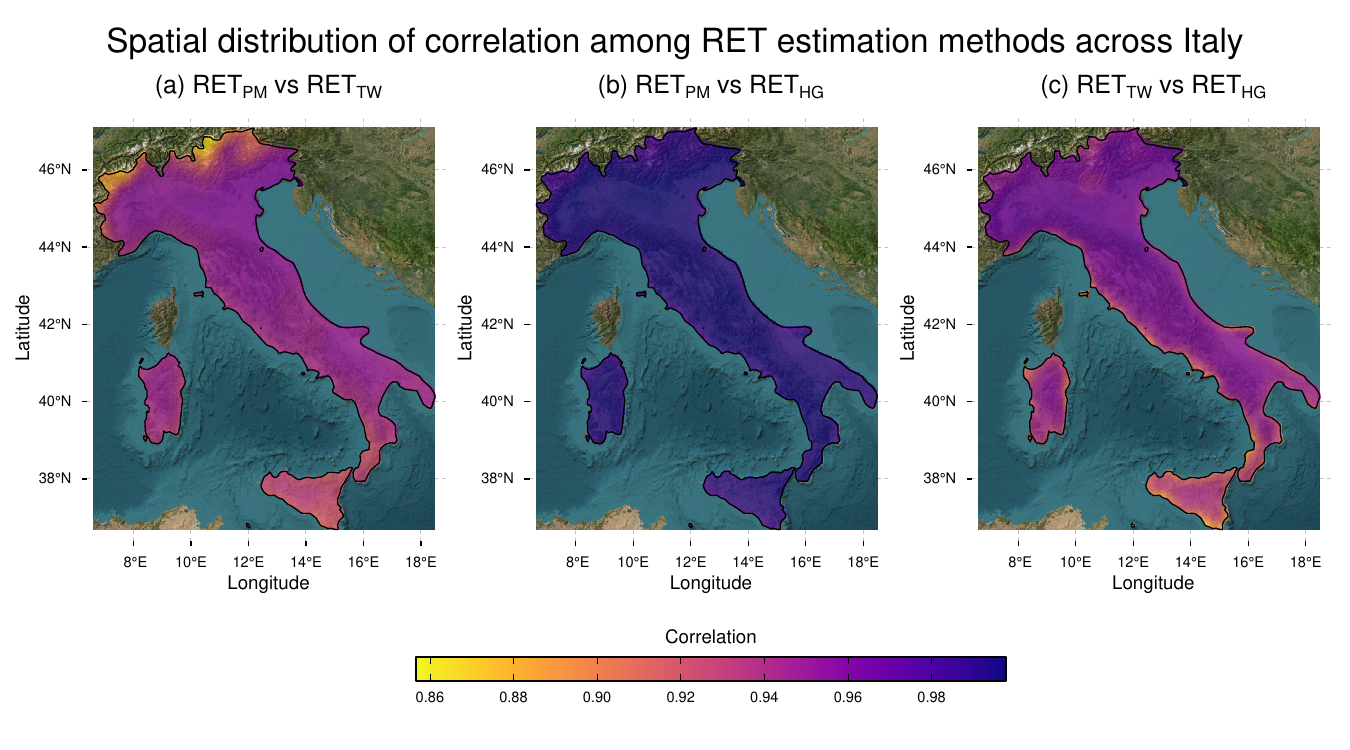}
    \caption{Spatial distribution of correlation  among the monthly
summed $\mathrm{RET}_{j}, j\in \{\mathrm{PM, TW, HG}\}$. }
    \label{fig:ret_cor_spatial}
\end{figure}

Figure~\ref{fig:ret_cor_spatial}(a--c) further illustrates the spatial distribution of correlation among the monthly summed $\mathrm{RET}_{j}$, $j \in {\mathrm{PM, TW, HG}}$. The correlation between $\mathrm{RET_{PM}}$ and $\mathrm{RET_{HG}}$ was consistently very high across Italy, ranging from 0.96 to 1.00, indicating an almost identical spatial and temporal variation between the two methods. In contrast, the correlations between $\mathrm{RET_{PM}}$ and $\mathrm{RET_{TW}}$ (0.85--0.94) and between $\mathrm{RET_{TW}}$ and $\mathrm{RET_{HG}}$ (0.88--0.94) were comparatively lower. The spatial distribution demonstrates that $\mathrm{RET_{PM}}$ and $\mathrm{RET_{HG}}$ capture highly similar RET dynamics, whereas $\mathrm{RET_{TW}}$ exhibits a comparatively different behavior due to its simplified temperature-driven formulation~\citep{lee2024sensitivity}.


\subsection{Distribution fitting to $\mathcal{D}_j$, $j\in \{\mathrm{PM, TW, HG}\}$ for SPEI}

\begin{figure}[t!]
  \centering

  \subfigure{
\includegraphics[width=0.30\textwidth]{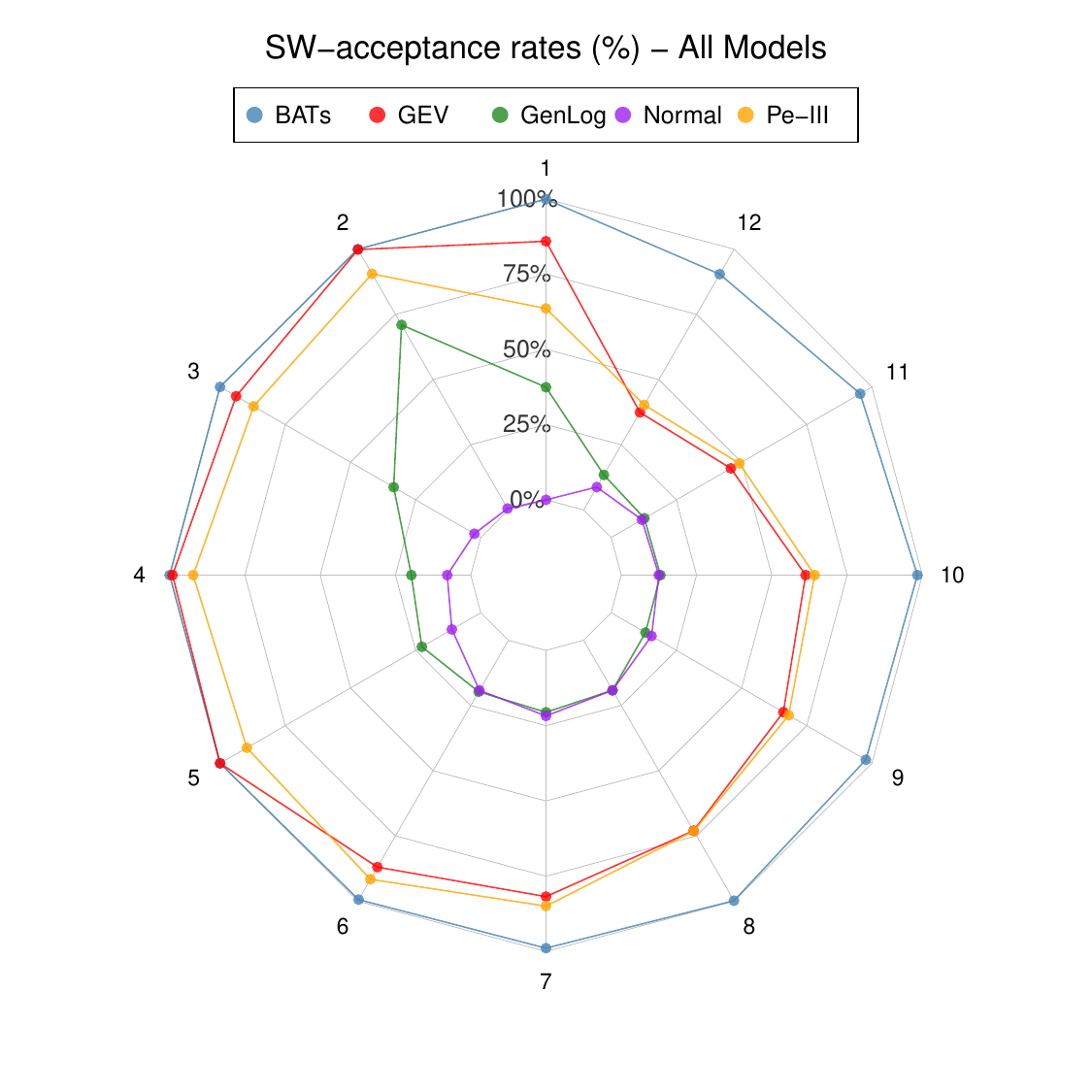}
  }
  \subfigure{
\includegraphics[width=0.30\textwidth]{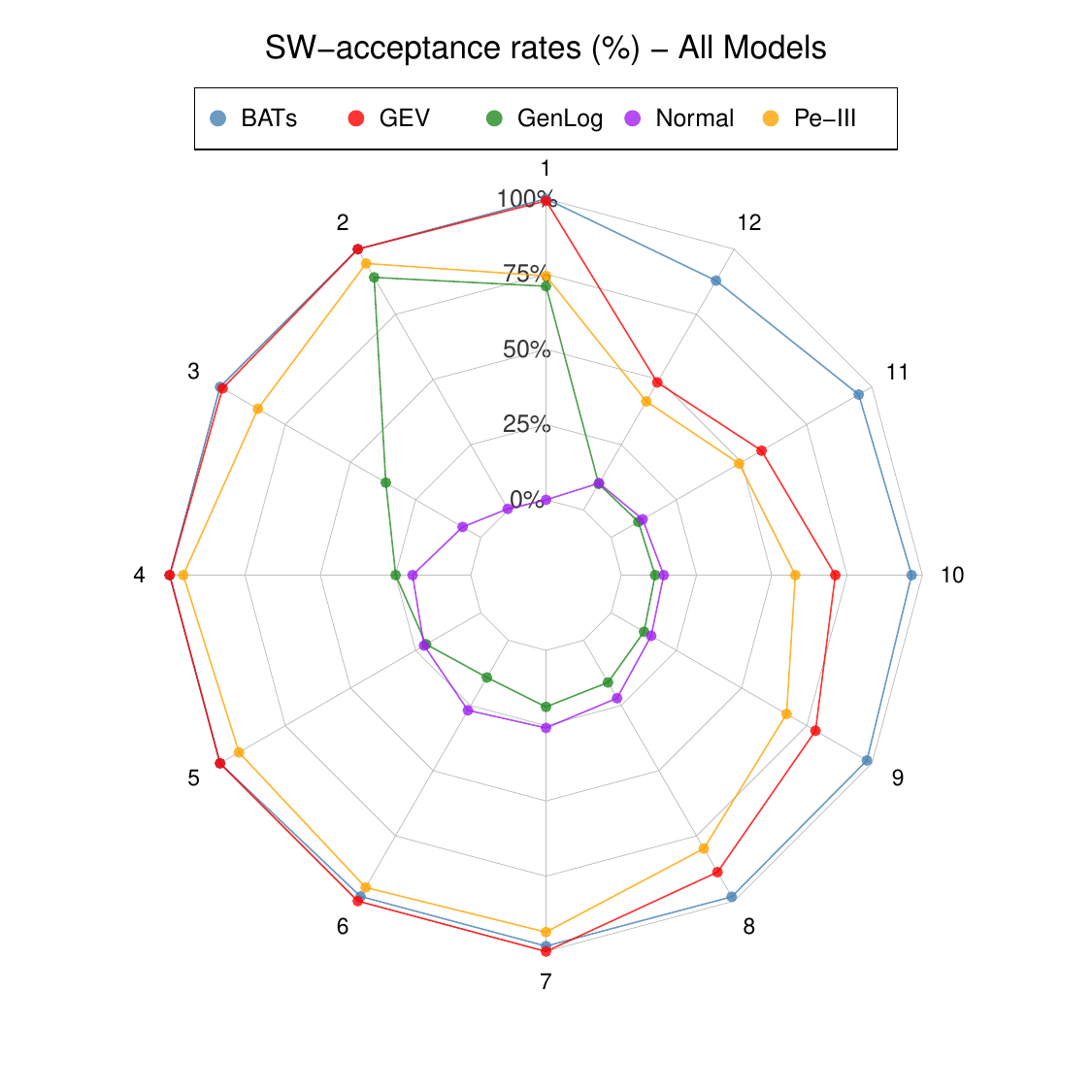}
  }
  \subfigure{
\includegraphics[width=0.30\textwidth]{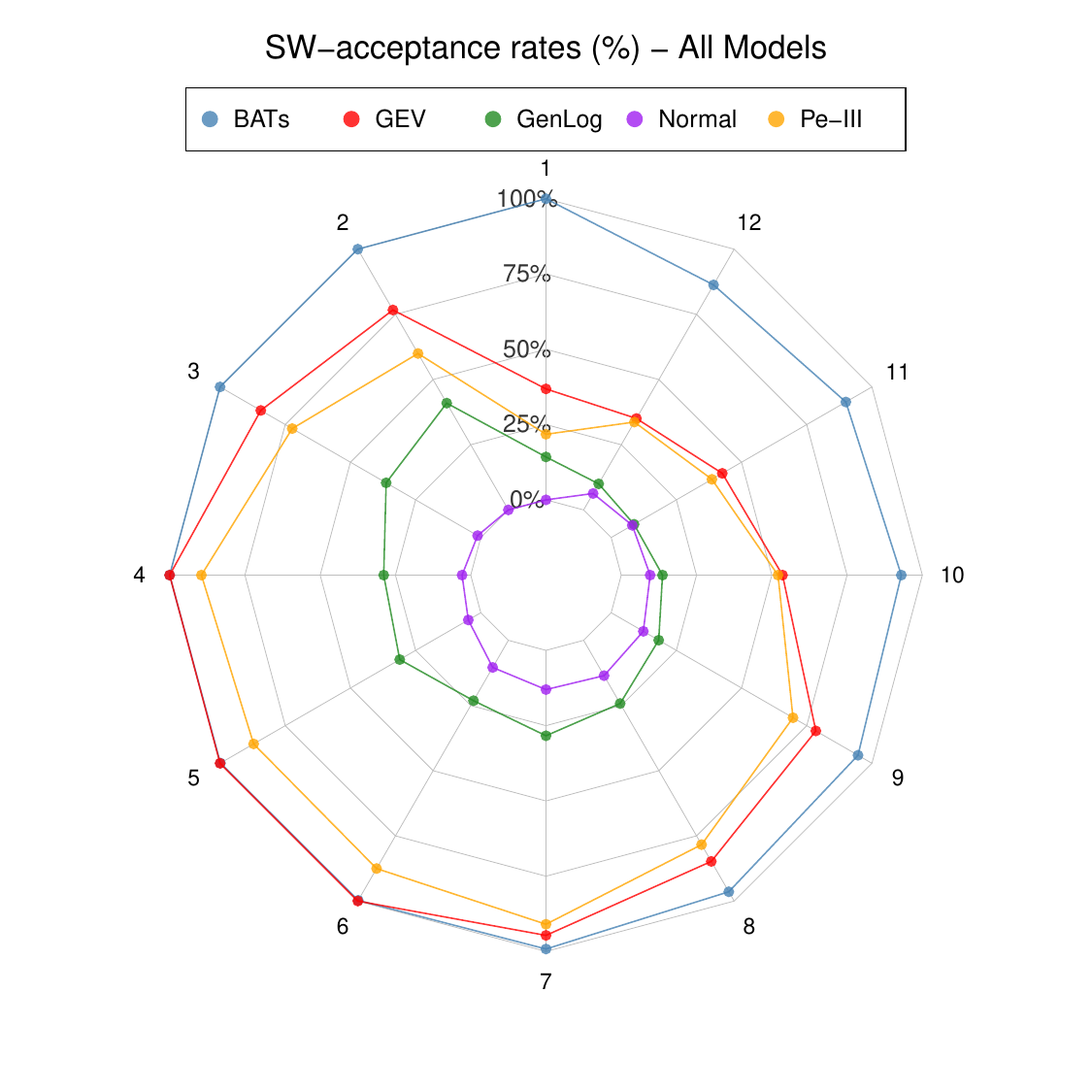}
  }
  \caption{Shapiro--Wilk normality test acceptance rate (\%) for SPEI obtained through different candidate distributions for time scales 1 to 12. Left to right: $\mathrm{RET_{PM}}$, $\mathrm{RET_{TW}}$, and $\mathrm{RET_{HG}}$ methods were used in $\mathcal{D}$.}
  \label{fig:sw_redar_all}
\end{figure}

Comparative analysis of candidate probability distributions fitted to $\mathcal{D}_j$, $j \in {\mathrm{PM, TW, HG}}$ at 730 grids across Italy for 1- to 12-month accumulation periods in SPEI demonstrates substantial differences in distribution suitability among the considered candidates. The SW normality test was employed for the evaluation, and the resulting acceptance rates based on adjusted p-values~\citep{benjamini1995controlling} for each distribution, RET method, and accumulation period are presented in Figure~\ref{fig:sw_redar_all}. The BATs distribution consistently outperformed the competing candidates across all RET methods and timescales, suggesting it better captures the statistical characteristics of $\mathcal{D}$ series under different climatic accumulation conditions.


For the PEM-based SPEI series~(Figure~\ref{fig:sw_redar_all}, left), the BATs distribution achieved exceptionally high acceptance rates across all timescales, starting at 99.9\% at the 1-month scale and remaining relatively stable even at longer accumulation periods, with the acceptance rate above 98.4\% at the 12-month timescale. This consistent dominance implies that the BATs distribution has considerable flexibility in modeling both short-term and long-term climatic fluctuations. In comparison, GEV and Pe-III appear closer to BATs on a 2- to 5-month timescale, with acceptance rates ranging from 85.89\% to 99.37\% and from 63.56\% to 92.19\%, respectively. Also, these distributions lose performance, with acceptance rates ranging from 86.98\% to 37.40\% and from 91.64\% to 40.27\% when the timescale increases to 6- to 12-month intervals, respectively. While both distributions performed satisfactorily for shorter accumulation periods, their acceptance frequencies progressively declined as timescales increased, indicating a diminished capacity to capture long-term SPEI dynamics. The GenLog distribution exhibited comparatively weak performance, particularly at medium and longer timescales, where acceptance rates mostly remained below 30\%. The normal distribution showed the poorest performance overall, with acceptance rates close to zero on shorter timescales and only marginal improvement on longer accumulation periods.

A similar pattern was observed for the TW-based SPEI series~(Figure~\ref{fig:sw_redar_all}, middle). The BATs distribution again emerged as the most suitable model, yielding acceptance rates from 100\% at the 1- to 5-month scale and remaining stable at 98.35\% to 87.94\% at the 6- to 12-month scale. The consistently high acceptance frequencies across all accumulation periods indicate that BATs provide a robust representation of the underlying SPEI distribution generated using the $\mathrm{RET_{TW}}$ method. The Pe-III and GEV distributions were the second-best alternatives, with acceptance rates ranging from 41.64\% to 95.47\% for Pe-III and from 51.09\% to 99.04\%  for GEV. Notably, both distributions showed relatively stronger performance on shorter timescales, particularly between 1 and 7 months, but their goodness of fit progressively deteriorated as accumulation periods increased. The GenLog distribution again demonstrated limited suitability, particularly beyond the 4-month timescale, where acceptance rates dropped below 20\%. The normal distribution remained the weakest candidate distribution, showing negligible acceptance at shorter timescales and consistently low goodness-of-fit across all accumulation periods.

For the HG-based SPEI series~(Figure~\ref{fig:sw_redar_all}, right), the BATs distribution maintained its superior performance relative to the other distributions. Acceptance rates ranged from 100\% to 99.18\% on the 1- to 7-month scale and remained stable at 96.43\% to 86.30\% on the 8- to 12-month scale, confirming the robustness and adaptability of BATs under the $\mathrm{RET_{HG}}$ framework. The Pe-III distribution showed moderate performance, with acceptance rates ranging from 33.69\% to 90.95\%, while the GEV distribution ranged from 35.07\% to 99.9\%. Both distributions displayed acceptable performance primarily at intermediate accumulation periods, although their fitting efficiency declined considerably at longer timescales. The GenLog and Normal distributions again exhibited substantially weaker performance. In particular, the Normal distribution produced near-zero acceptance rates at shorter timescales. It remained below 13\% across all accumulation periods, indicating its inability to adequately capture the non-normal characteristics and the variability structure of the SPEI data.

\begin{figure}[t]
  \centering
  \subfigure{
\includegraphics[width=1\textwidth, height=0.8\textwidth]{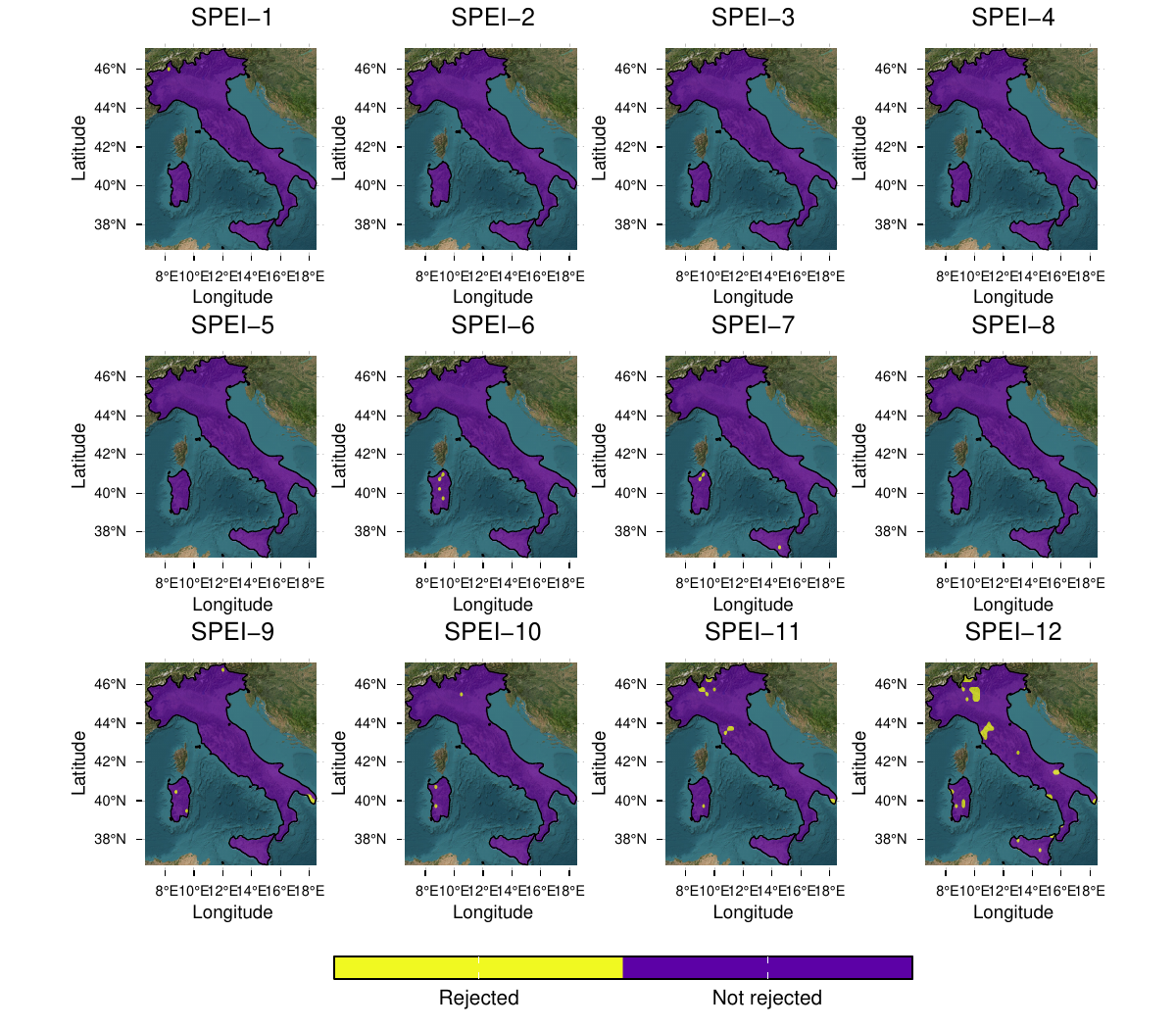}
 }
  \caption{Spatial comparison of the SW test at each grid cell under $\mathrm{RET_{PM}}$ for BATs-SPEI-1 through BATs-SPEI-12. }
  \label{fig:SW_BAT_grid}
\end{figure}

We also provided additional goodness-of-fit assessments using the Anderson–Darling test and Kolmogorov–Smirnov–based Lilliefors test in 
Figures~\textcolor{red}{S.2} and~\textcolor{red}{S.3} to further distinguish among the candidate distributions. Both tests exhibit patterns that are fully consistent with the SW test and provide additional evidence in favor of the BATs fitting superiority. A detailed previous study by~\citep{stagge2015candidate} identified the GEV distribution as a preferred choice for SPEI applications in Europe. While GEV provides a reasonably good fit for extreme values, BATs offer much greater flexibility and can adjust both lower- and upper-tail behaviors simultaneously. This added flexibility is especially crucial for drought indices, where asymmetry, heavy-tailed behavior, and varying extreme conditions commonly arise across different climate regions. Consequently, BATs can also handle a broader range of distributional shapes while preserving stability in the central Student t-distribution. Grid-wise Shapiro–Wilk test results for BATs at different accumulation scales (1 to 12) under the $\mathrm{RET_{PM}}$ method are presented in Figure~\ref{fig:SW_BAT_grid}, while the corresponding results for the remaining candidate distributions are provided in the Supplementary Material (Figures~\textcolor{red}{S.4}--\textcolor{red}{S.7}).
The individual grid acceptance patterns for the SW test highlight that the BATs fit better across a wide range of climatic zones than the competing models.

\begin{figure}
  \centering
  \subfigure{
\includegraphics[width=1\textwidth, height=0.5\textwidth]{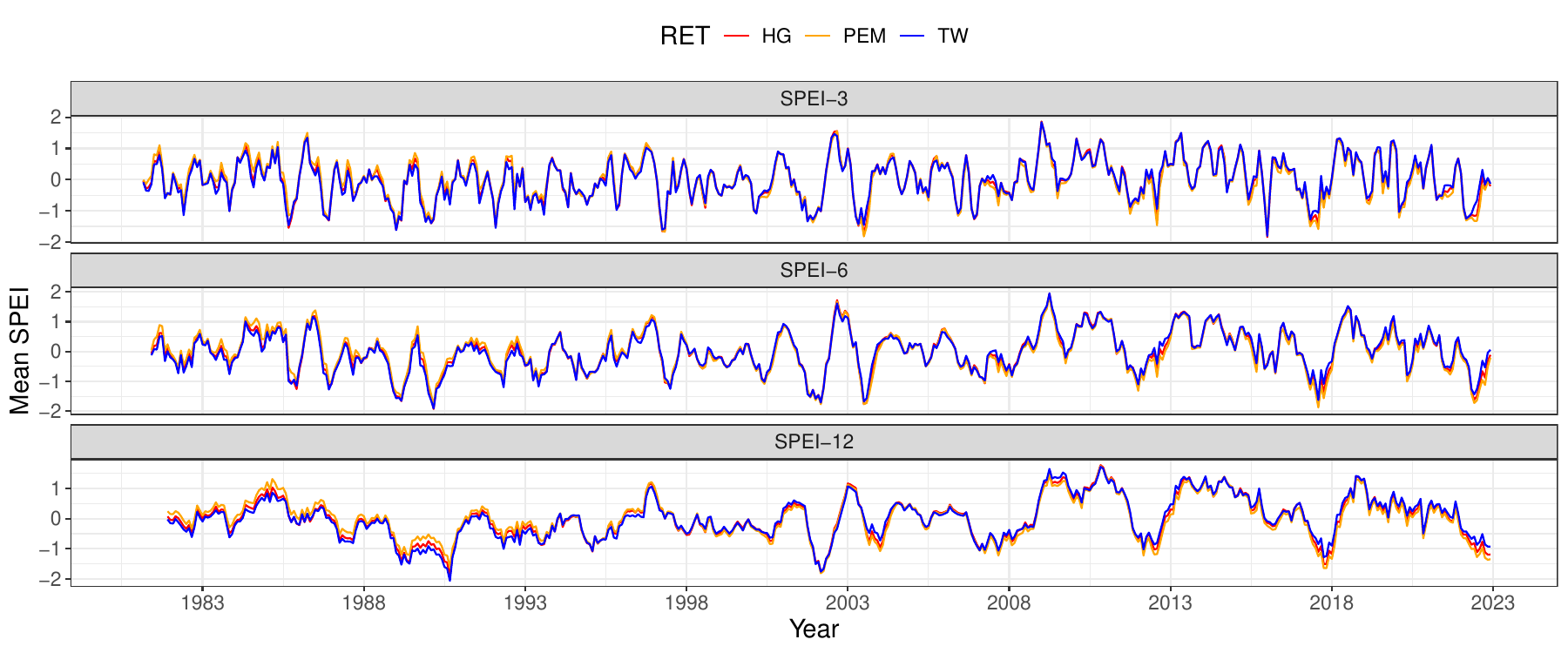}
 }
  \caption{Temporal variation of spatially averaged BATs-SPEI time series over Italy calculated at 3-, 6-, and 12-month accumulation scales based on the  $\mathrm{RET_{PM}}$, $\mathrm{RET_{TW}}$, and $\mathrm{RET_{HG}}$.  }
  \label{fig:SPEI_mean}
\end{figure}

The temporal evolution of the spatially averaged SPEI-3, SPEI-6 and SPEI-12  calculated via BATs models (Figure~\ref{fig:SPEI_mean}) reveals a high degree of consistency among the  $\mathrm{RET}_j$, $j\in \{\mathrm{PM, TW, HG}\}$ formulations, as evidenced by the near-complete overlap of the corresponding curves. This indicates that although the three methods differ in their estimates of atmospheric evaporative demand, these differences have a negligible influence on the resulting standardized SPEI values when estimated via BATs. Consequently, the overall drought signal and the identification of wet and dry periods remain largely unaffected by the choice of RET method. 
Indeed, as shown by earlier work~\citep{lee2024sensitivity}, the SPEI is relatively insensitive to the choice of PET formulations used in its computation at regional and national levels, whereas $\mathrm{RET_{PM}}$ is considered more practical~\citep{stagge2015candidate}.

The SPEI-3 series exhibits the largest temporal variability, characterized by rapid oscillations between positive and negative values that reflect short-term meteorological fluctuations and seasonal moisture anomalies. As the accumulation timescale is extended to 6 and 12 months, the series grows increasingly smooth, reflecting the integration of moisture conditions over longer durations. The SPEI-6 series captures seasonal-to-interannual drought persistence, whereas SPEI-12 emphasizes long-term hydroclimatic variability and sustained drought episodes. This progressive decline in variability with increasing timescales is an inherent characteristic of multiscalar drought indices and has been extensively reported in drought monitoring research~\citep{pascale2025widespread}.

Although several periods of below-normal and above-normal moisture conditions can be identified throughout the record, the overall spatially averaged SPEI will remain close to zero. This behavior is expected because SPEI is a standardized index with a long-term mean near zero. Moreover, averaging across all grid cells tends to offset regional wet and dry anomalies over Italy.  Consequently, the relatively steady overall mean should not be interpreted as the absence of drought; rather, it indicates the simultaneous presence of contrasting hydroclimatic conditions across different areas of Italy. Broadly speaking, the temporal evolution of drought conditions in Italy is primarily driven by climatic variability rather than by the specific evapotranspiration scheme adopted, whereas longer accumulation periods increasingly emphasize persistent drought behavior.

\subsection{Drought trends and long-range dependences across Italy}
\begin{figure}[t]
  \centering
  \subfigure{
\includegraphics[width=1\textwidth, height=0.5\textwidth]{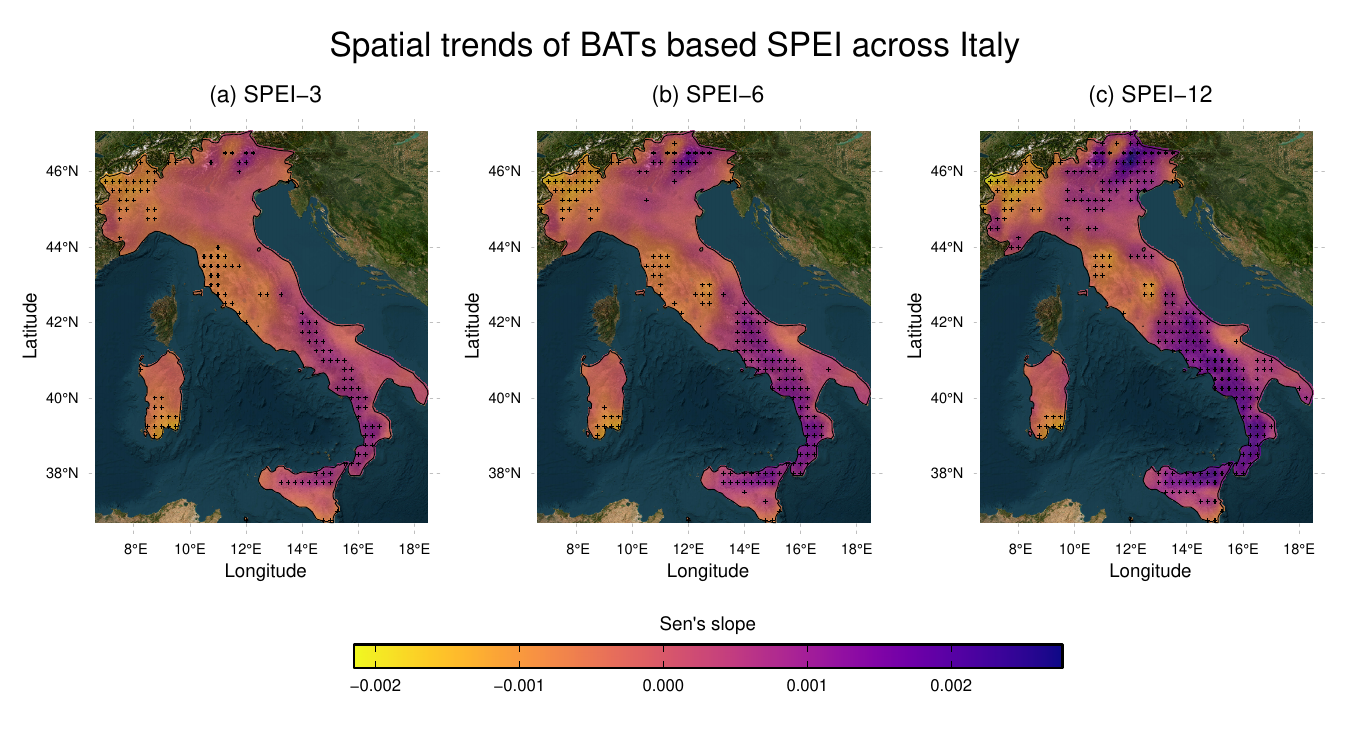}
  }
  \caption{Spatial distribution of trends for BATs-SPEI-3, SPEI-6 and SPEI-12 under $\mathrm{RET_{PM}}$.  }
  \label{fig:trend_BAT}
\end{figure}

As noted, the BATs model maintained its superior
performance for all RET equations. Therefore, we estimated drought trends at different time scales for SPEI calculated from BATs for $\mathrm{RET_{PM}}$. 
The spatial distribution of SPEI-3, SPEI-6, and SPEI-12 month timescale trends reveals significant regional differences across Italy. In Figure~\ref{fig:trend_BAT}, positive values characterize increasing wet conditions, while negative slope values correspond to precipitation deficit and highlight increasing drought conditions.

In a short period (e.g., SPEI-3), positive and statistically significant trends prevail in some areas of southern Italy, i.e., Campania, Calabria, Basilicata, and northern Sicily, indicating a gradual increase in short-term soil wetness. At the same time, some parts of northwestern Italy, especially in Piedmont and Liguria, Tuscany, and southwestern Sardinia, show significant negative trends, indicating that these areas are becoming drier due to more frequent short-term droughts. Central Italian regions have mostly neutral or very weak trends.

A more contrasting picture emerges for SPEI-6, where southern and southeastern Italian regions dominate in terms of positive trends, with a high share of statistically significant values, i.e., increasing moisture availability over time at these timescales. Negative trends persist in parts of northwestern Italy and the Alpine region, particularly in southeastern Apulia and southwestern Sardinia, indicating that the medium-term moisture deficiency in these regions is worsening. Central regions tend to have slight trends.


The SPEI-12 shows significant differences between northern and southern Italy. In large areas of southern Italy, significant positive trends are observed, whereas in Sicily the opposite is evident, with negative trends characterized by increasing moisture deficits and drought conditions. In contrast, the northwestern and northeastern parts of Italy are characterized by increasing moisture availability and decreasing drought conditions.


\begin{figure}[t]
  \centering
  \subfigure{
\includegraphics[width=1\textwidth, height=0.35\textwidth]{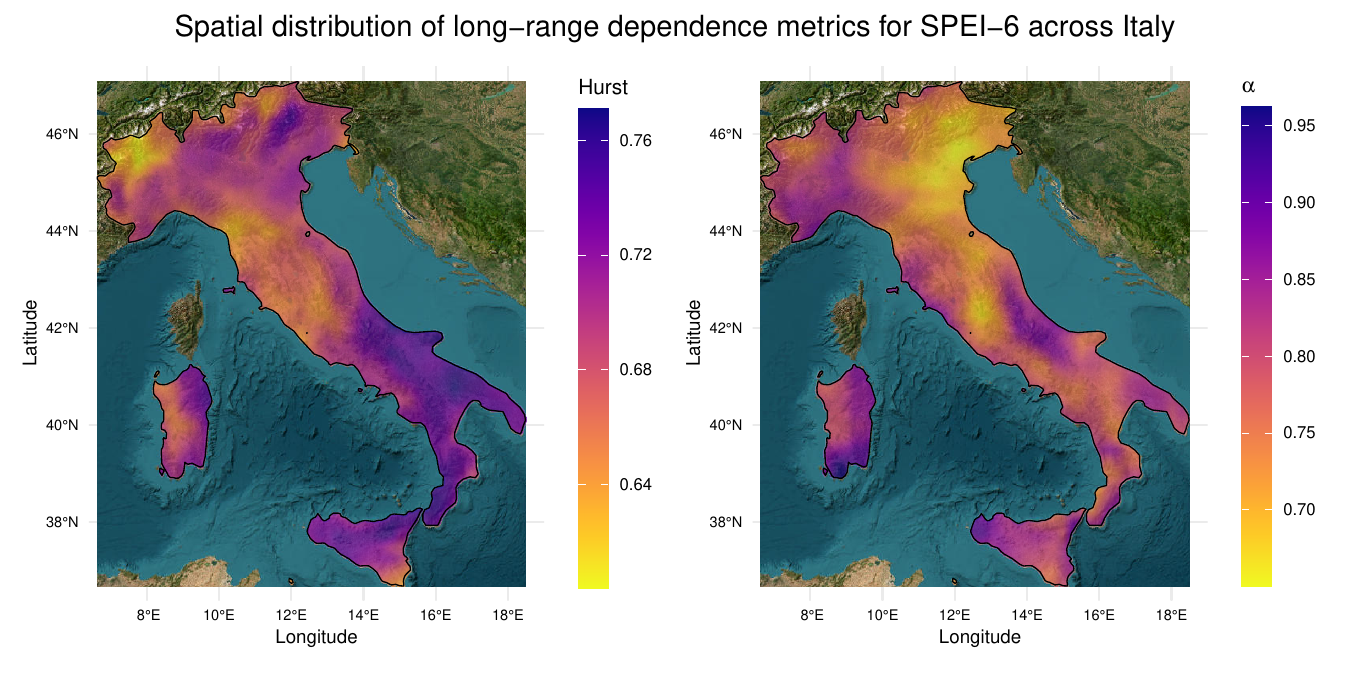}
  }
  \caption{Spatial distribution of long-range dependence and scaling metrics: (left) Hurst exponent; (right) detrended fluctuation analysis exponent for BATs-SPEI-6 under $\mathrm{RET_{PM}}$. }
  \label{fig:dep_long}
\end{figure}

The spatial pattern of the Hurst exponent ($\mathcal{H}$) for SPEI-6 across Italy (Figure~\ref{fig:dep_long} left) shows a strong meridional gradient in the long-term memory behavior. The largest $\mathcal{H}$ values, typically ranging from 0.73 to 0.81, are observed in the southern peninsular regions and major islands, including extensive areas of Calabria, Sicily, and Sardinia, contrary to the general expectation that persistence is higher at high latitudes. Our results also support the findings presented in~\citet{GRANATA2026134428}. These high values of $\mathcal{H}$ indicate strong long-term persistence, which means that both dry and wet spells have a tendency to last for long durations. Such behavior is typical of hydroclimatic systems driven by low-frequency variability and high autocorrelation and is probably due to a combination of long seasonal patterns, reduced synoptic disturbance, and the moderating effect of the adjacent Mediterranean Sea.

Towards the north, $\mathcal{H}$ values tend to decrease, reaching moderate values (0.65-0.71) in central Italy and along the Tyrrhenian coastal zone. The lowest values (0.59–0.65) are found in the Alpine foothills, the northeastern Veneto–Friuli Plain and parts of Tuscany. In these northern areas, the SPEI-6 signal is characterized by an increased frequency of regime shifts and shorter anomaly durations, indicative of the effects of complex orography, increased frontal activity, and a more dynamic synoptic environment. In these regions, hydroclimatic anomalies are more temporally fragmented, as the memory of past events is more readily disrupted by extratropical disturbances and mesoscale convective systems.

Any value of  $\mathcal{H}>0.5$  indicates persistence. However, it could be noted that the magnitude contains meaningful physical differences in drought memory. Moderate persistence ($\mathcal{H}\approx 0.59-0.65$) means that anomalies are concentrated over relatively short, often intraseasonal time scales. Intermediate values ($\mathcal{H}\approx 0.66-0.74$) represent stronger anomaly persistence that can cross seasonal boundaries, increasing the chance of extended droughts. High persistence $\mathcal{H}>0.80$  indicates a very slow decay of autocorrelation and indicates multi-year drought development and a greatly diminished ability to recover rapidly to normal conditions.  These distinctions are not merely statistical calculations; they have tangible implications for drought duration, spread, and predictability~\citep{GRANATA2026134428}.

The complementary DFA exponent $(\alpha)$ (Figure~\ref{fig:dep_long} right) confirms and sharpens this spatial interpretation. Across Italy, DFA values range from 0.61 to 0.97, affirming the presence of persistent, scale-invariant temporal correlations. Values from 0.61 to 0.85 observed along the Alpine margins and northeastern lowlands correspond to moderate persistence and relatively high-frequency variability, consistent with fast hydroclimatic fluctuations and limited drought memory. Conversely, DFA exponents exceeding 0.85, found in southern Italy, eastern Sicily, northeastern Apulia, and Sardinia, indicate very strong persistence, reflecting multi-year cycles and pronounced climatic inertia that prolong drought conditions. In DFA theory, the scaling exponent  $\alpha$  can theoretically fall between 0 and 2: for stationary fractional Gaussian noise, $0< \alpha< 1$, with  $\alpha = 0.5$ indicating white noise and  $\alpha > 0.5$ signaling persistent correlations; for non-stationary fractional Brownian motion, $1 < \alpha < 2$ \citep{GRANATA2026134428}.
Together, these findings underscore strong spatial heterogeneity in the temporal structure of hydroclimatic variability across Italy. Southern and island regions exhibit inertia-driven memory regimes, whereas northern areas are more prone to rapid oscillations and to external forcing, often linked to passing synoptic systems originating over continental Europe. This spatial divergence has important implications for drought forecasting and water management: in southern Italy, the persistence of anomalies may enable more reliable long-term predictions, whereas the transient character of northern regimes demands adaptive, high-frequency monitoring. Ultimately, the combined interpretation of Hurst and DFA metrics provides a nuanced, spatially continuous view of drought persistence, moving beyond simple duration-based assessments to offer a rigorous foundation for region-specific modeling and early warning system design under evolving climatic conditions.

\subsubsection{Spatial distribution of the persistence}

Drought patterns vary substantially across Italy due to diverse regional differences in climate, topography, and hydro-meteorological conditions~\citep{GRANATA2026134428}. Therefore, both spatial and nonspatial versions of the model proposed in Section~\ref{Bayesian_model} are fitted and compared. Model predictive performance was evaluated using Leave-One-Out Cross-Validation Information Criterion~\citep{vehtari2015loo}: $\mathrm{LOOIC} = -2\mathrm{ELPD_{LOO}} $ with expected log
predictive density $\mathrm{ELPD_{LOO}}= \sum_{i=1}^n \log p(y_i | y_{-i}) $, where  $y_{-i}$ the dataset used for the model is the observation $i$ removed, and the Watanabe-Akaike information criterion~\citep{watanabe2010asymptotic}: $\mathrm{WAIC}=-2
\left(
\mathrm{lppd}
-
p_{\mathrm{WAIC}}
\right),$ where
$
\mathrm{lppd}
=
\sum_{i=1}^{n}
\log
\left[
\int p(y_i \mid \theta)
p(\theta \mid y)
\, d\theta
\right]
$
is the log pointwise predictive density, and
$
p_{\mathrm{WAIC}}
=
\sum_{i=1}^{n}
\mathrm{Var}_{\theta|y}
\left[
\log p(y_i \mid \theta)
\right]
$
is the effective number of parameters. Smaller LOOIC and WAIC values indicate better predictive performance of the model. 

Table~\ref{tab:loo_waic} summarizes the model predictive performance results. The spatial Gaussian process model achieved a higher $\mathrm{ELPD_{LOO}}=-36923.0$, compared with $\mathrm{ELPD_{LOO}}=-37069.7$ the non-spatial model. Consistent with this result, the spatial model also yielded lower values of both LOOIC and WAIC than the non-spatial counterpart.
The agreement between $\mathrm{ELPD_{LOO}}$, $\mathrm{LOOIC}$, and $\mathrm{WAIC}$ suggests that incorporating spatial heterogeneity successfully captures key spatial patterns in drought recovery dynamics. 
The improved predictive performance of the spatial model provides strong evidence that spatial dependence is a key component of the recovery process and should be incorporated into drought recovery modeling. We further investigate the model performance using posterior predictive checks and residual spatial diagnostics (Figures~\textcolor{red}{S.8}--\textcolor{red}{S.9}).
The checks showed that both the non-spatial and spatial process models successfully reproduced the marginal distribution of the observed response, as evidenced by close agreement between observed and replicated values across the entire range of the response variable. This indicates that both models provide an adequate fit to the observed data distribution.
Despite their similar predictive performance with respect to the response distribution, the two models differed substantially in their treatment of spatial dependence. The Moran's scatterplot of residuals~\citep{shortridge2007practical} from the non-spatial model demonstrated a significant positive relationship between residuals and their spatial lag, corresponding to a Moran's I statistic of 0.4222 ($p=6.61×10^{-315}$
). This strong residual spatial autocorrelation indicates that important spatial structure remained unexplained. In contrast, the residuals from the spatial process model showed no apparent spatial structure, with Moran's I near zero (-0.0007) and a non-significant p-value (0.4757), suggesting that the residuals are spatially independent. Hence, we consider only the spatial model in the rest of the analysis.


\begin{table}[htbp]
\centering
\caption{Comparison of predictive performance between the non-spatial and spatial Gaussian process models.}
\begin{tabular}{lccc}
\hline
Model & ELPD$_{\mathrm{LOO}}$ & LOOIC & WAIC \\
\hline
Non-spatial & -37069.7 & 74139.3 & 74139.3 \\
Spatial GP  & \textbf{-36923.0 }&\textbf{73846.0} & \textbf{73845.9} \\
\hline
\end{tabular}
\label{tab:loo_waic}
\end{table}

The spatial field of the total slope parameter \(\beta_{\text{total}}(s) = [\beta_0 + \beta(s)\)] (Figure~\ref{fig:persistance} a) constitutes the primary inferential output of the Bayesian spatial BMCD model. Because \(\beta_{\text{total}}(s)\) determines the sign and magnitude of duration dependence in the logit hazard, its spatial pattern provides an integrated view of drought regime type unavailable from classical frequency analyses. A negative value indicates that recovery probability declines with elapsed duration; a value near zero implies approximate memorylessness; a positive value would indicate self-terminating behavior.

The mapped spatial field reveals a significant north–south variability across the Italian peninsula and major islands. Across the Po Plain, Lombardy, and the southern Alpine fringe, \(\beta_{\text{total}}(s)\) approaches zero, indicating approximately geometric spell durations consistent with a first-order Markov chain. Moving southward through the central Apennines, values become progressively more negative with the gradient becoming steeper below approximately \(40^\circ\)N. In the southern Apennines, Calabria, Sicily, and southern Sardinia, \(\beta_{\text{total}}(s)\) reaches its most negative values (often \(<-0.6\))  characterizing strongly persistent regimes where recovery hazard declines steeply with duration.

The physical mechanisms underpinning this heterogeneity are well established. Southern Italy lies within the core Mediterranean climate zone, where summer precipitation is virtually absent, evapotranspiration greatly exceeds precipitation for four to six months annually, and interannual variability is modulated by the Azores High and North African anticyclone~\citep{ferrari2013influence}. Once initiated, drought activates positive feedback: reduced soil moisture suppresses evapotranspiration and boundary-layer moistening, diminishing convective precipitation \citep{seneviratne2010investigating}; increased sensible heat flux warms and dries the lower troposphere, reinforcing subsidence \citep{kim20242022}; and persistent anticyclonic blocking inhibits Atlantic moisture advection \citep{lionello2018relation}. The net effect is a self-reinforcing drought state, which is precisely the interpretation of negative \(\beta_{\text{total}}(s)\).

Northern Italy and the Po Valley experience a more continental regime with evenly distributed annual precipitation, stronger frontal and/or orographic mechanisms, and weaker land–atmosphere coupling \citep{giorgi2008climate}. As a result, drought termination is less dependent on elapsed duration because precipitation events can be generated by synoptic disturbances independent of the land surface state. The near-zero \(\beta_{\text{total}}(s)\) values in the north are therefore consistent with a regime driven primarily by large-scale circulation rather than local feedbacks~\citep{kim2025characterizing}. By contrast, the Alpine fringe presents the least negative (and locally slightly positive) values, consistent with high precipitation, orographic uplift, and strong Atlantic influence \citep{rettig2024responses}. Slight positivity in some high-elevation cells may reflect reduced drought frequency and wider posterior intervals.

The spatial continuity of the \(\beta_{\text{total}}(s)\) field is of interpretive significance. The Matérn 5/2 GP prior induces smooth interpolation between stations, producing a continuous gradient rather than artificial discontinuities~\citep{porcu2024matern}. This continuity reflects the physical expectation that persistence mechanisms transition gradually across climate gradients. Posterior variance is higher in sparser areas, providing honest quantification of spatial interpolation uncertainty.

This pattern extends previous station-based analyses. \citet{spinoni2014world} documented longer and more severe drought events in southern Europe using SPI-based indices, while \citet{peres2023dynamic} identified higher drought recurrence intervals in Sicily and Calabria. Our model provides a mechanistic explanation: droughts are not merely more frequent or severe in the south; their termination dynamics are fundamentally different, characterized by a declining hazard that traps the system in the drought state once a sufficient duration has been established.

\subsubsection{Recovery and Survival Probability Curves}

\begin{figure}[t]
\centering
\subfigure[]{
\includegraphics[
  width=0.45\textwidth,
  height=0.25\textheight
]{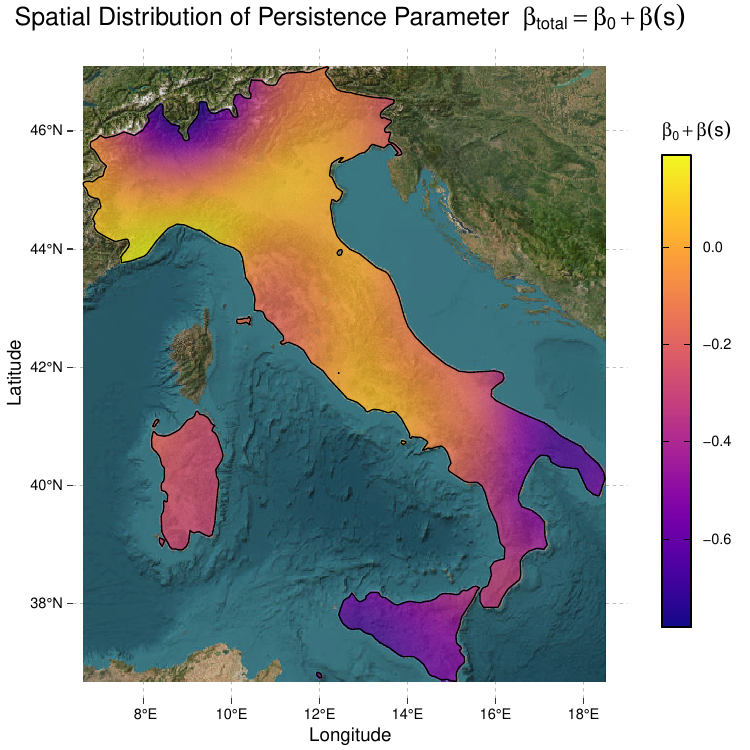}
}
\hfill
\subfigure[]{
\includegraphics[
  width=0.45\textwidth,
  height=0.25\textheight
]{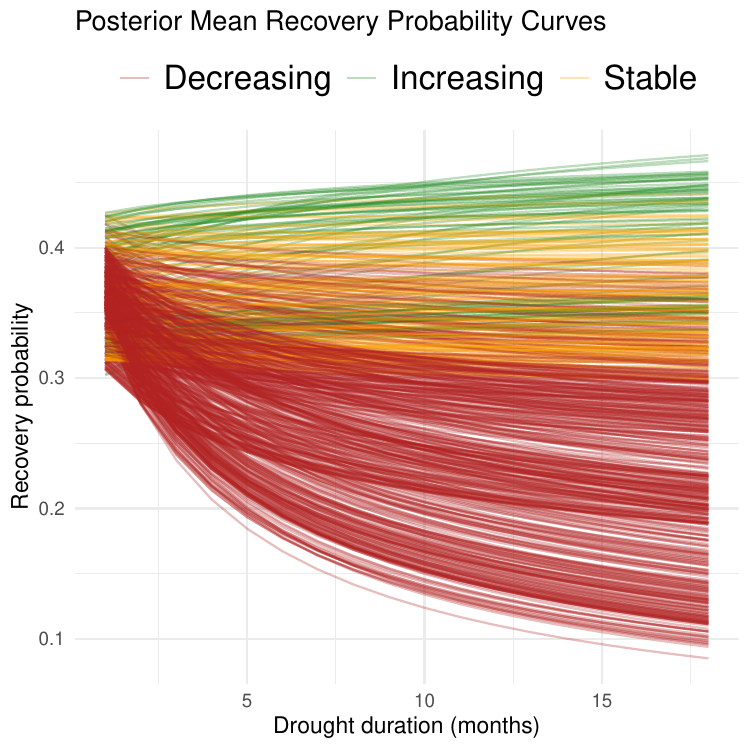}
}
\vspace{0.3cm}

\subfigure[]{
\includegraphics[
  width=\textwidth,
  height=0.25\textheight
]{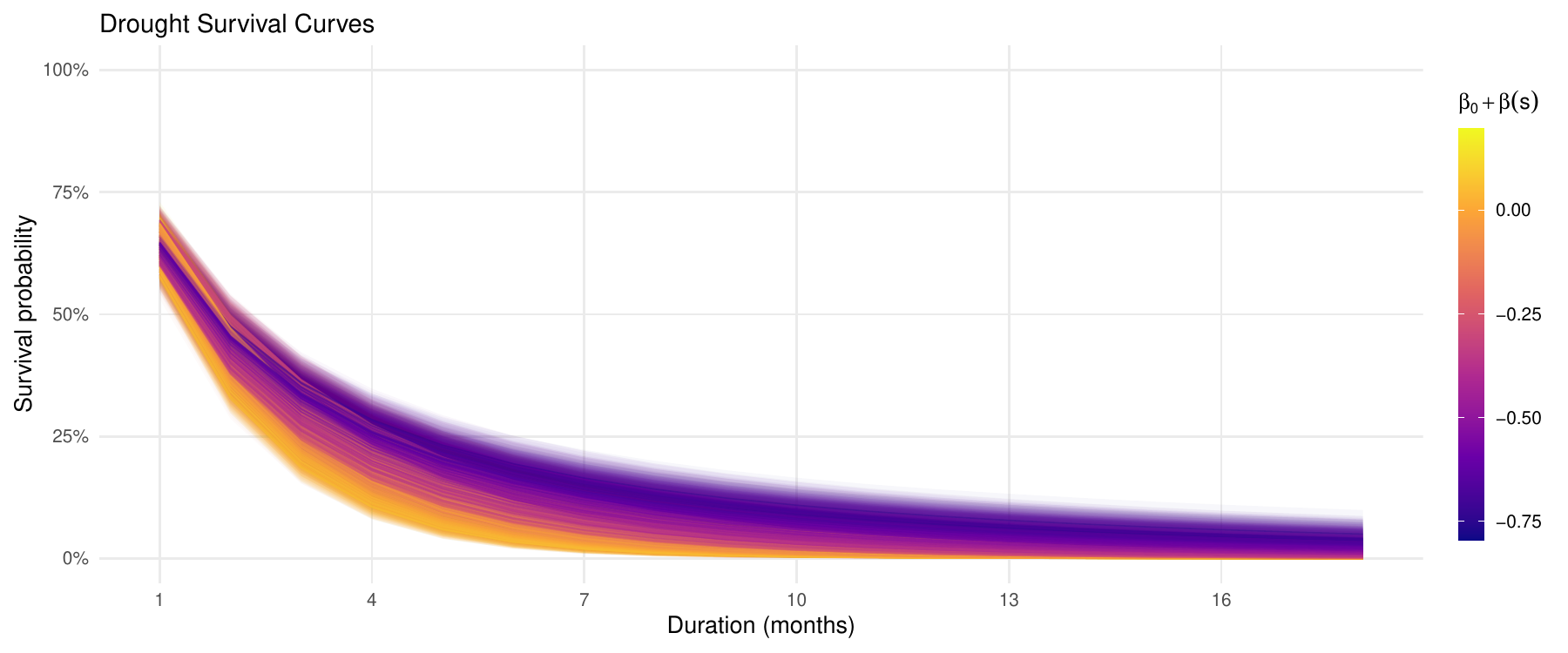}
}

\caption{Spatial and temporal patterns of drought persistence across Italy. (a) Spatial distribution of the total duration-dependence parameter, $\beta_{\mathrm{total}}(s)$. (b) Posterior mean drought recovery probability curves. (c) Corresponding drought survival probability curves.
}
\label{fig:persistance}
\end{figure}

The posterior mean recovery probability curves (Figure~\ref{fig:persistance}, b), stratified by persistence regime, translate the spatial $\beta_{\text{total}}(s) $ field into a temporal perspective. For locations with strong negative $\beta_{\text{total}}(s) $ values (southern Italy, Sicily, and Sardinia), the recovery curve declines monotonically with a characteristic convex shape. At $(d = 1)$ month, the recovery probability is relatively high $(0.30-0.45)$, but it falls rapidly to $(0.10-0.20)$ by $(d = 6) $ months and to negligible values $(<0.05)$ by $(d = 12)$ months. This 'drought trap' mechanism~\citep{zscheischler2020typology} shows that duration dependence in the south is a dominant feature of termination dynamics rather than a marginal statistical effect. Under the BMCD alternating renewal equivalence \citep{doize2026}, the declining hazard generates heavy-tailed spell-duration distributions; locations with strong negative $\beta_{\text{total}}(s) $ carry substantially more probability mass at durations exceeding 12 months than a geometric distribution would imply. Risk assessments based on memoryless models will therefore systematically underestimate the probability of multi-year drought.

Locations in the stable regime (the Po Plain and central Apennines) display approximately flat recovery curves $(0.15-0.25)$ consistent with geometric spell durations. In this regime, the processes that produce precipitation operate largely independently of land-surface conditions, so the likelihood of termination does not vary systematically with event duration~\citep{taszarek2021global}. By contrast, the increasing regime, in which recovery probability rises with duration, is less common; it appears in parts of the northern Alpine fringe and may reflect soil moisture reaching a depletion floor or simply be a statistical artifact of sparse data.
The posterior survival curves (Figure~\ref{fig:persistance}, c) represent the probability that a drought has not terminated by month $d$. Curves for southern locations (strongly negative $\beta_{\text{total}}(s))$ decline slowly and exhibit heavily right-skewed tails at $(d = 12)$ months; survival probabilities of (0.15–0.25) are common, and at $(d = 18)$ months many locations retain non-trivial mass above $(0.05-0.10)$. In other words, once a drought in southern Italy lasts a full year, there is a substantial chance (15–25\%) that it will continue for at least another six months. In contrast, the survival curves for the north decline quickly, falling below 0.10 by $(d = 6)$ months and approaching zero by $(d = 12)$ months, consistent with the geometric decay pattern implied by a constant hazard.

\section{Conclusion}
\label{sec:conclusion}
This study aims to enhance the characterization and modeling of drought persistence over a heterogeneous region by introducing a spatially duration-augmented framework for drought assessment. To achieve this objective, we also address two key methodological challenges that directly influence drought persistence estimates: (i) the estimation of reference evapotranspiration (RET), which determines the climatic water balance used in drought monitoring, and (ii) the sensitivity of probability distributions employed in the SPEI.  Using the high-resolution MADIA dataset from Italy, we systematically evaluate these components and integrate them into a unified framework for representing the spatial and temporal persistence of drought conditions.

The main results highlight the following: First, the choice of RET formulation has only a marginal effect on the drought signal at the national scale. Although the $\mathrm{RET_{TW}}$ diverged most from the closely correlated $\mathrm{RET_{PM}}$ and $\mathrm{RET_{HG}}$ fields, consistent with its simplified temperature-based form, the spatially averaged SPEI series at the 3-, 6-, and 12-month scales (Figure~\ref{fig:SPEI_mean}) were almost indistinguishable across all three methods. This confirms that the SPEI is not particularly sensitive to the RET model and supports the continued use of $\mathrm{RET_{PM}}$ as a practical reference, as well as supports the claim made by~\citet{stagge2015candidate} regarding the use of $\mathrm{RET_{PM}}$.

Second, the BATs model consistently outperformed existing candidates across all RET methods and accumulation scales. Under $\mathrm{RET_{PM}}$, the Shapiro–Wilk test achieved acceptance rates above 99.9\% for BATs at the one-month scale and remained near 98.4\% at twelve months, with comparable dominance under the other two formulations; the Pearson Type III and GEV distributions were the closest competitors but degraded sharply at longer scales, while the GenLog and Normal distributions proved inadequate. Flexible tails allow BATs to capture the asymmetry and heavy-tailed behavior of the climatic water-balance series far more faithfully than existing alternatives, establishing it as a new candidate distribution for SPEI construction in heterogeneous climates.

Third, the Bayesian spatial duration-augmented Markov-chain framework described recovery dynamics more accurately and comprehensively than its non-spatial counterpart, as confirmed by improved leave-one-out and information-criterion scores (Table~\ref{tab:loo_waic}). The fitted model revealed a significant south-to-north variation in temporal memory that mirrors the independent Hurst and DFA diagnostics. Southern peninsular and other regions (e.g., Calabria, Sicily, and Sardinia) exhibit strong persistence and a duration-dependent hazard that declines as droughts lengthen, making prolonged events progressively harder to terminate. Northern and Alpine regimes, by contrast, show near-zero duration dependence consistent with rapidly fluctuating, circulation-driven conditions~\citep{manco2025identifying}. The practical implication is that the conventional duration-blind SPEI classification is adequate in the north but systematically underestimates the risk of extended drought in the south, where a persistence-aware, duration-conditioned monitoring scheme is needed. By embedding duration dependence into Gaussian process priors, the proposed framework produces spatially continuous, uncertainty-quantified persistence fields, thereby providing a formal statistical characterization of the previously mentioned spatial variability in drought persistence across Italy~\citep{bischof2025exploring}.


These contributions advance both methodology and practice in drought monitoring. Methodologically, we introduce a flexible distribution and a spatial persistence model that better reflect the statistical behavior of drought than commonly used tools. Practically, our results show thatdistributional and persistence assumptions, rather than the RET formulation, most strongly shape drought characterization across Italy. Although our analysis focuses on Italy, the framework is modular and transferable and can be applied to other Mediterranean and semi-arid regions facing intensifying climatic pressures.

\section*{Funding}
This work has received funding from the European Union's Horizon Europe research and innovation program under the Marie Sklodowska-Curie grant agreement No. 101126636. The authors also acknowledge the support of the Research Council of Norway through its Centre of Excellence Integreat – The Norwegian Centre for Knowledge-driven Machine Learning, project number 332645.







\bibliographystyle{apalike}  
 
\begin{center}
    \bibliography{References}
\end{center}



\newpage





\renewcommand{\thetable}{S.\arabic{table}}
\setcounter{table}{0}
\renewcommand{\thefigure}{S.\arabic{figure}}
\setcounter{figure}{0}

\renewcommand{\theequation}{S.\arabic{equation}}
\setcounter{equation}{0}

\renewcommand{\thesection}{S.\arabic{section}}
\setcounter{section}{0}

\setcounter{page}{1}
\resetlinenumber[1] 
\begin{center}
{\LARGE\bfseries Supplementary Material}\\[1em]

{\large for}\\[0.5em]
{\LARGE\bfseries A spatial duration-augmented framework for drought persistence}\\[1.5em]

Touqeer Ahmad$^{1}$ and Thordis Linda Thorarinsdottir$^{1}$\\[0.5em]

$^{1}$ Department of Mathematics, University of Oslo, P.O. Box 1053 Blindern, 0316 Oslo, Norway.
\end{center}

\vspace{1cm}
\section{Additional Details Regarding Methodology}\label{add_method}
\subsection{BATs candidate distribution}\label{add_Bats}

Let $\mathcal{D}_{j}, j\in \{\mathrm{PM, TW, HG}\}$ be a deficit series aggregated according to a chosen time scale; the  BATs model is specified by its cumulative distribution function (CDF) as
\begin{equation}\label{bat-cdf}
    \mathcal{F}_{\theta}(\mathcal{D}_{j}) = \mathcal{T}_\nu(\mathcal{H}_\theta(\mathcal{D}_{j})), \quad j\in \{\mathrm{PM, TW, HG}\}
\end{equation}
where $\mathcal{T}_\nu$ is the CDF of the Student's $t$ distribution with $\nu > 0$ degrees of freedom, and $\mathcal{H}_\theta(\mathcal{D}_{j})$ is a monotonic transformation determined by the parameter vector $\boldsymbol{\theta}$. To define a proper model, we suppose that a strictly increasing function \( \Upsilon: \mathbb{R} \rightarrow \mathbb{R} \), which satisfies:\[
\lim_{y \to -\infty} \Upsilon(\mathcal{D}_{j}) = 0, \quad \text{and} \quad \lim_{x \to \infty} \left[\Upsilon(\mathcal{D}_{j}) - \mathcal{D}_{j}\right] = 0. \] By using cumulative distribution function $\mathcal{G}$ with an analytic, positive density and finite mean, \citet{stein2021parametric} define $\Upsilon$ function which satisfying the above limiting conditions as 
$\Upsilon(x) = \int_{-\infty}^y \mathcal{G}(x)dx.$ 
Adopting the standard logistic distribution $
\mathcal{G}(y) = {e^y}/{1 + e^y},$ we get $\Upsilon(y) = \log(1 + e^y)$ and define as a CDF $\mathcal{H}_\theta(\mathcal{D}_{j})$ as
\begin{equation}\label{inner-cdf}
    \mathcal{H}_\theta(\mathcal{D}_{j}) = \left[1 + \gamma_2 \Upsilon\left(\frac{\mathcal{D}_{j} - \alpha_2}{\beta_2}\right)\right]^{1/\gamma_2} - \left[1 + \gamma_1 \Upsilon\left(\frac{\alpha_1 - \mathcal{D}_{j}}{\beta_1}\right)\right]^{1/\gamma_1},
\end{equation}
where $\theta = (\alpha_i \in \mathbb{R}, \beta_i>0, \gamma_i \in \mathbb{R})$ are the  location, scale, shape parameters
for $i = 1$ (lower tail) and $i = 2$ (upper tail). In analogy with the other extreme value distributions, the tail index parameter $\gamma_i, i\in(1,2)$ determines the tail's behavior: positive values indicate a heavy-tailed distribution with infinite support, whereas negative values correspond to a light-tailed distribution with bounded support in the tail. 

Differentiating the expression \eqref{bat-cdf} yields the probability density function of the distribution as
\begin{equation}\label{dens-bat}
    f(\mathcal{D}_{j}|\theta) = t_\nu(\mathcal{H}_\theta(\mathcal{D}_{j})) \cdot \mathcal{H}_\theta'(\mathcal{D}_{j}),
\end{equation}

\section{Additional Results}
\subsection{Spatial distribution of $\mathrm{RET}$ differences}
\begin{figure}[h]
\includegraphics[width=1\textwidth,height=8cm]{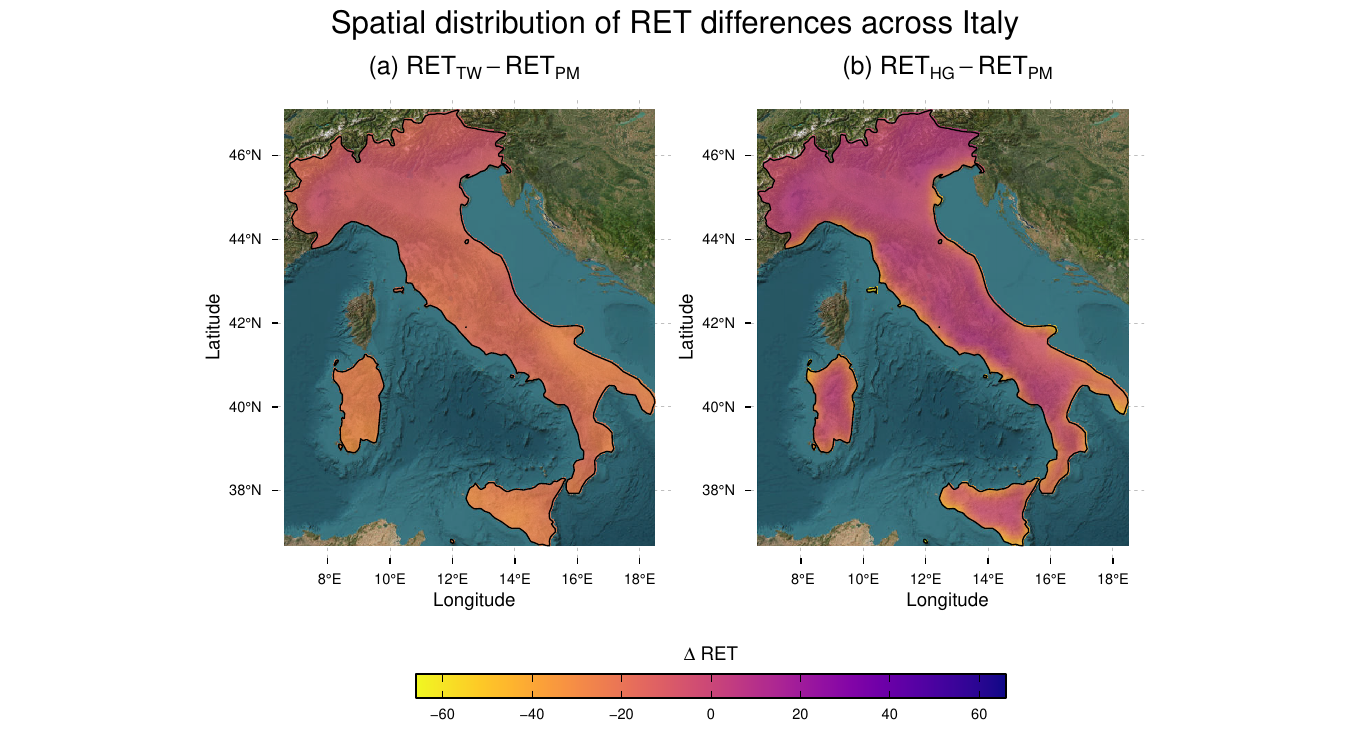}
    \caption{Spatial distribution of differences of $\mathrm{RET}_{j}$, $j \in {\mathrm{PM, TW, HG}}$ at one-month time scale.}
    \label{fig:ret_difference_S}
\end{figure}

As discussed in the main document, the overall spatial pattern was broadly consistent among the three methods, yet notable differences in RET magnitude and spatial variability were also evident. The $\mathrm{RET_{PM}}$ and $\mathrm{RET_{HG}}$ methods both showed greater spatial heterogeneity and generally higher RET values over most of Italy, with the most pronounced discrepancies in the southern regions and along the coasts. Conversely, $\mathrm{RET_{TW}}$ yielded lower $\mathrm{RET}$ estimates and exhibited smoother spatial gradients, particularly across the Alpine and semi-arid areas. Figure~\ref{fig:ret_difference_S} depicts the spatial distribution of the differences between $\mathrm{RET_{TW}}$ and $\mathrm{RET_{HG}}$ relative to $\mathrm{RET_{PM}}$. The $\mathrm{RET_{TW}}$ method appears to underestimate RET compared with $\mathrm{RET_{PM}}$, as its temperature-driven formulation, which does not explicitly incorporate other atmospheric drivers such as humidity, radiation, and wind speed~\textcolor{red}{\citep{stagge2015candidate}}. In contrast, $\mathrm{RET_{HG}}$ exhibits slightly smaller deviations in coastal zones and in the Puglia region.

\subsection{Distribution fitting to $\mathcal{D}_j$, $j\in \{\mathrm{PM, TW, HG}\}$ for SPEI}

\begin{figure}
  \centering
  \subfigure{
\includegraphics[width=0.30\textwidth]{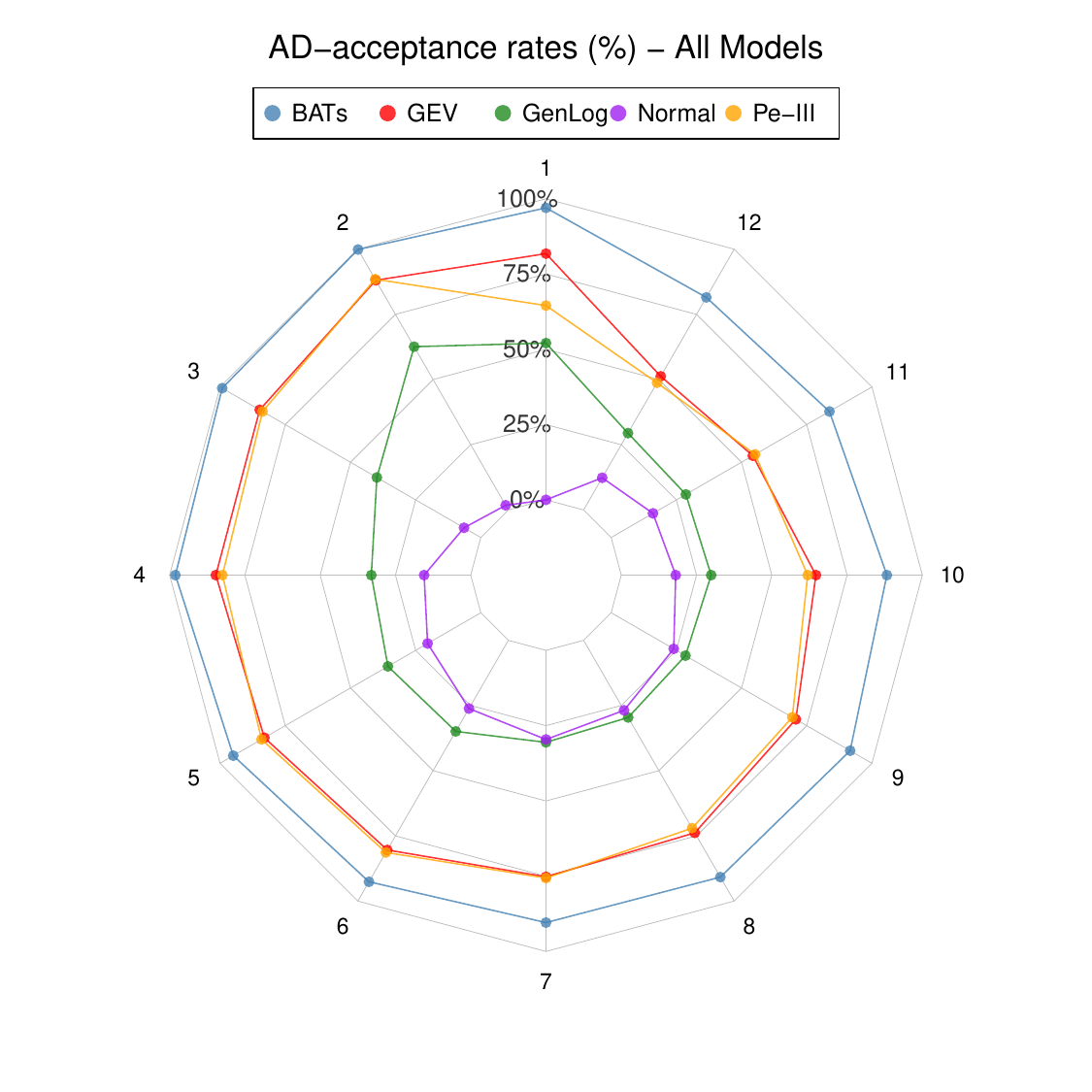}
  }
  \subfigure{
\includegraphics[width=0.30\textwidth]{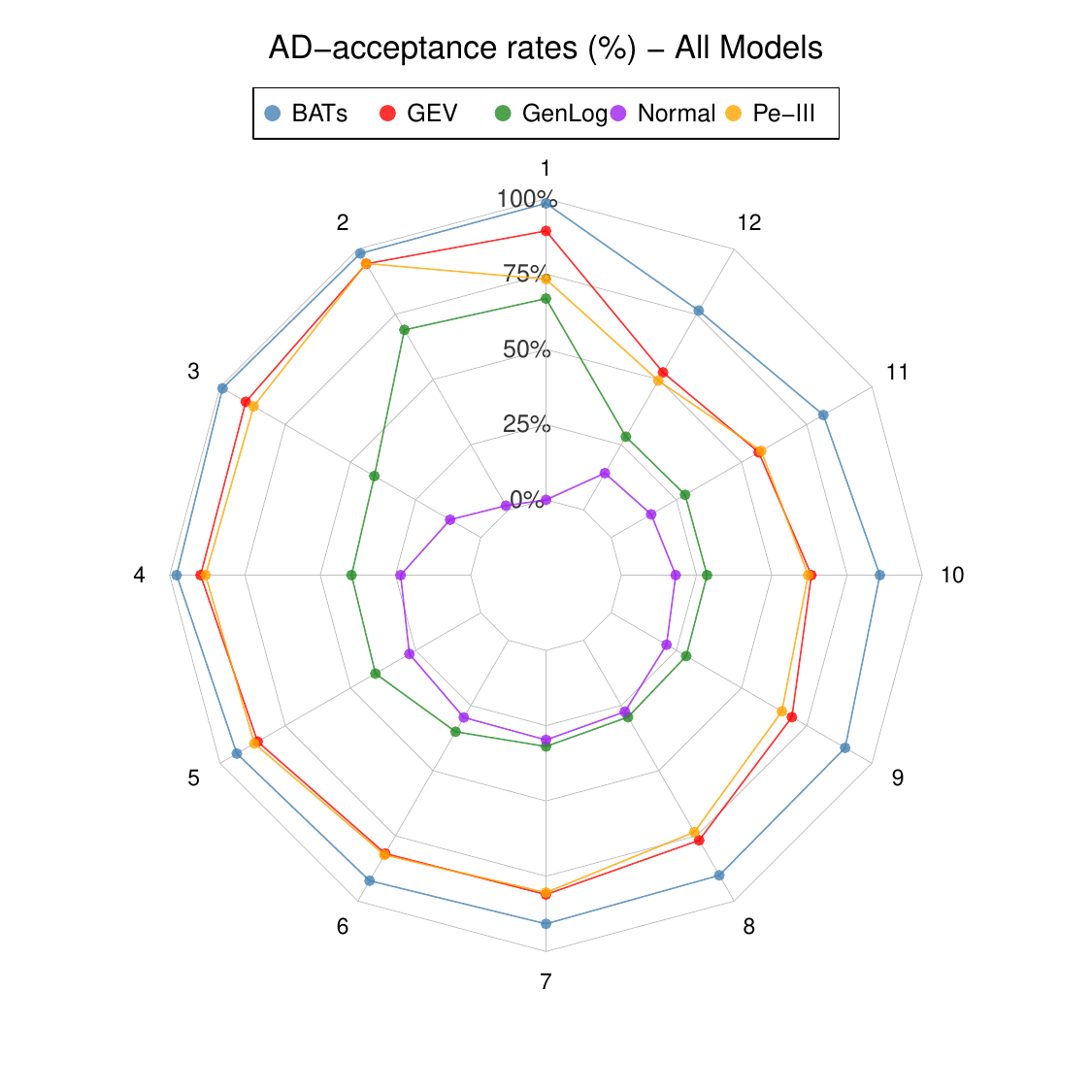}  }
  \subfigure{
\includegraphics[width=0.30\textwidth]{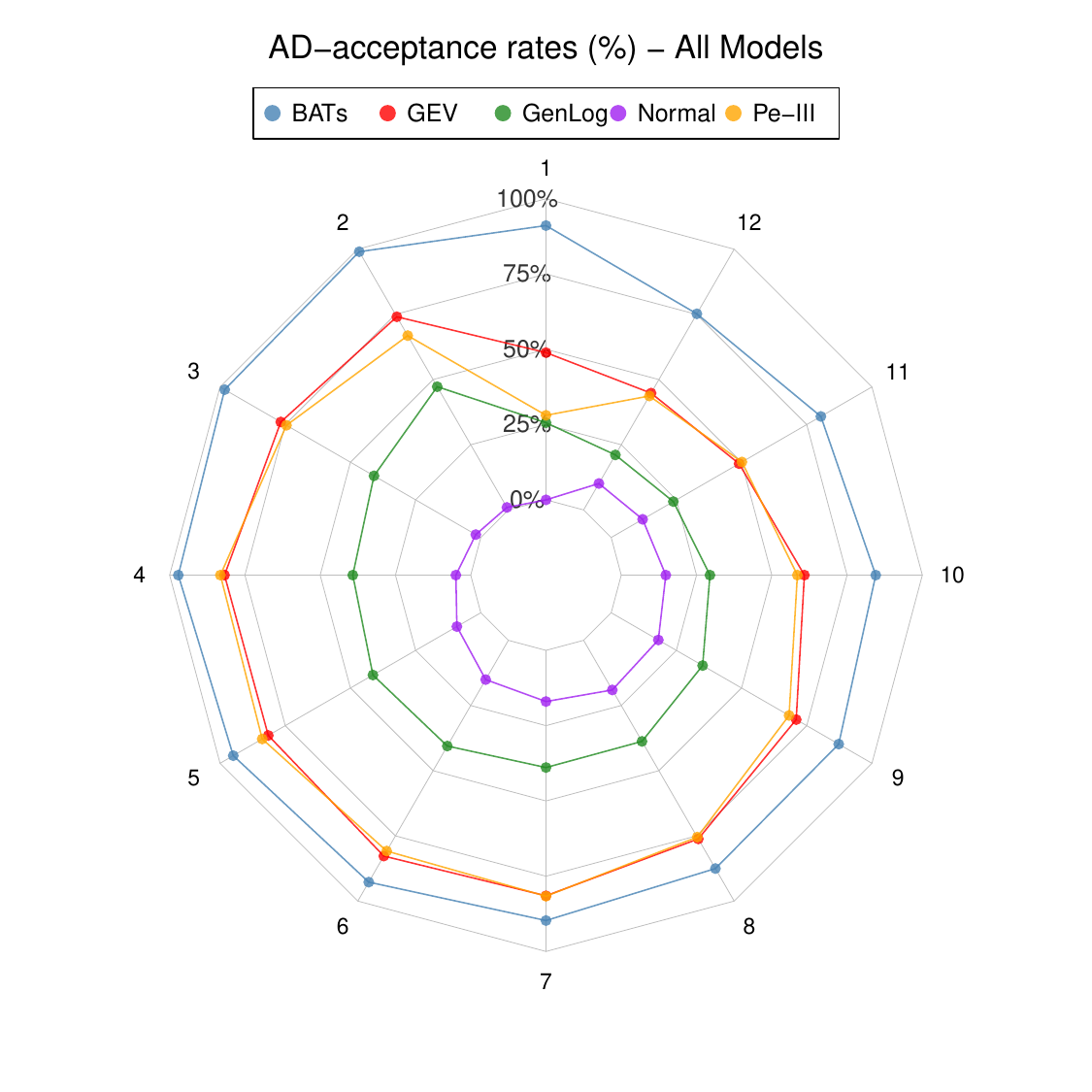}
  }
  \caption{Anderson-Darling normality test acceptance rate (\%) for SPEI obtained through different candidates of distributions for time scales 1 to 12. Left to right: $\mathrm{RET_{PM}}$, $\mathrm{RET_{TW}}$, and $\mathrm{RET_{HG}}$ methods were used in~$\mathcal{D}$.}
  \label{fig:AD_redar_all}
\end{figure}

\begin{figure}
  \centering
  \subfigure{
\includegraphics[width=0.30\textwidth]{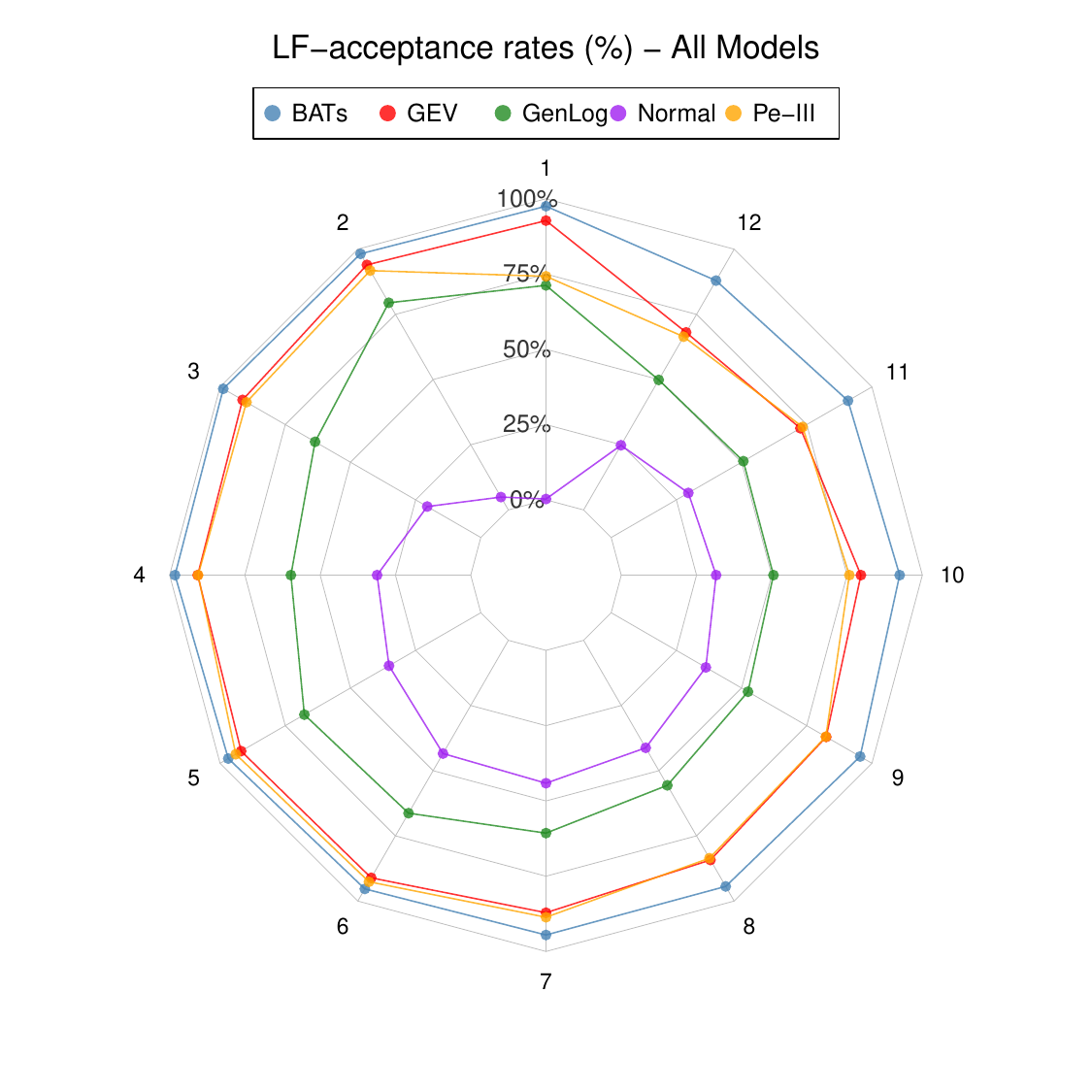}
  }
  \subfigure{
\includegraphics[width=0.30\textwidth]{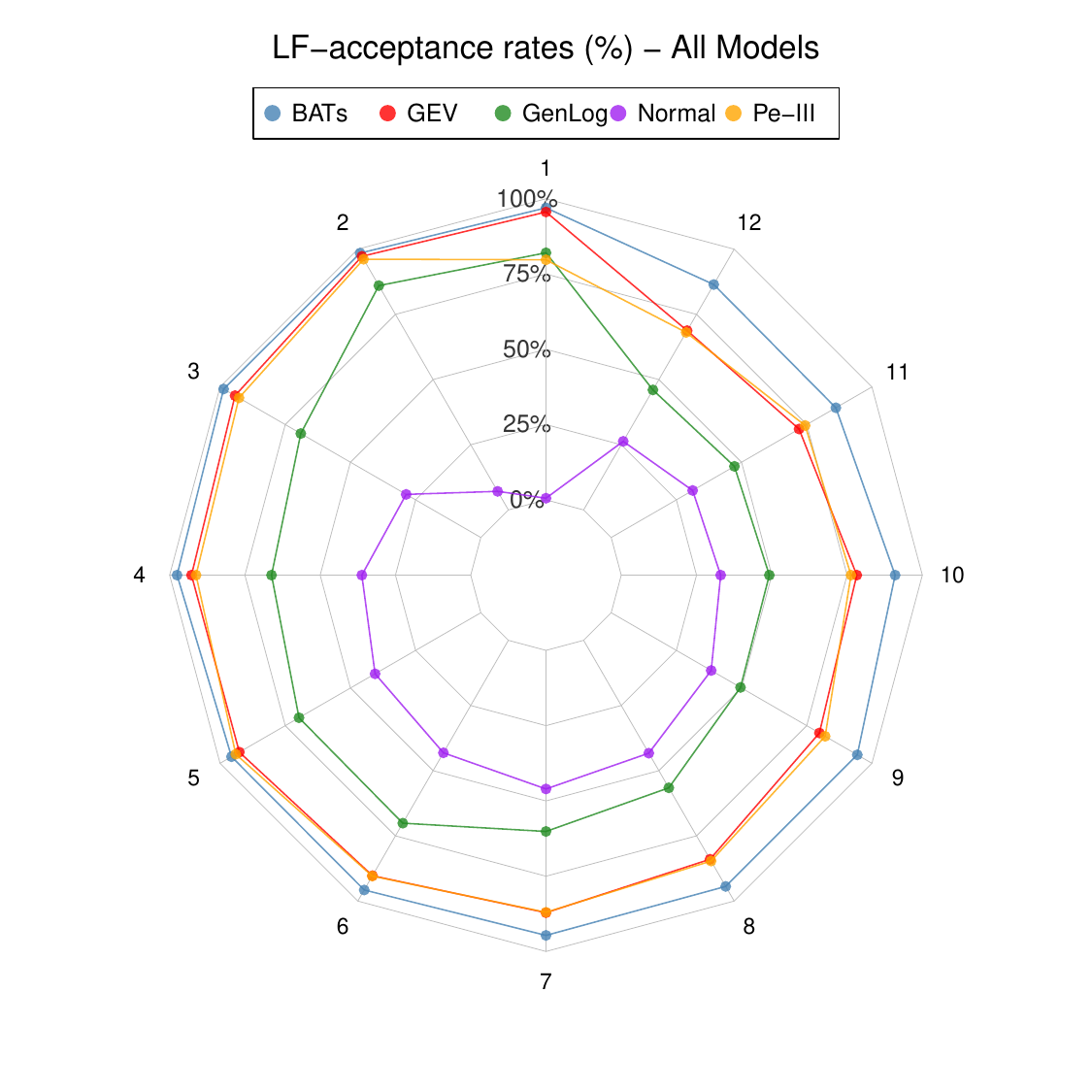}  }
  \subfigure{
\includegraphics[width=0.30\textwidth]{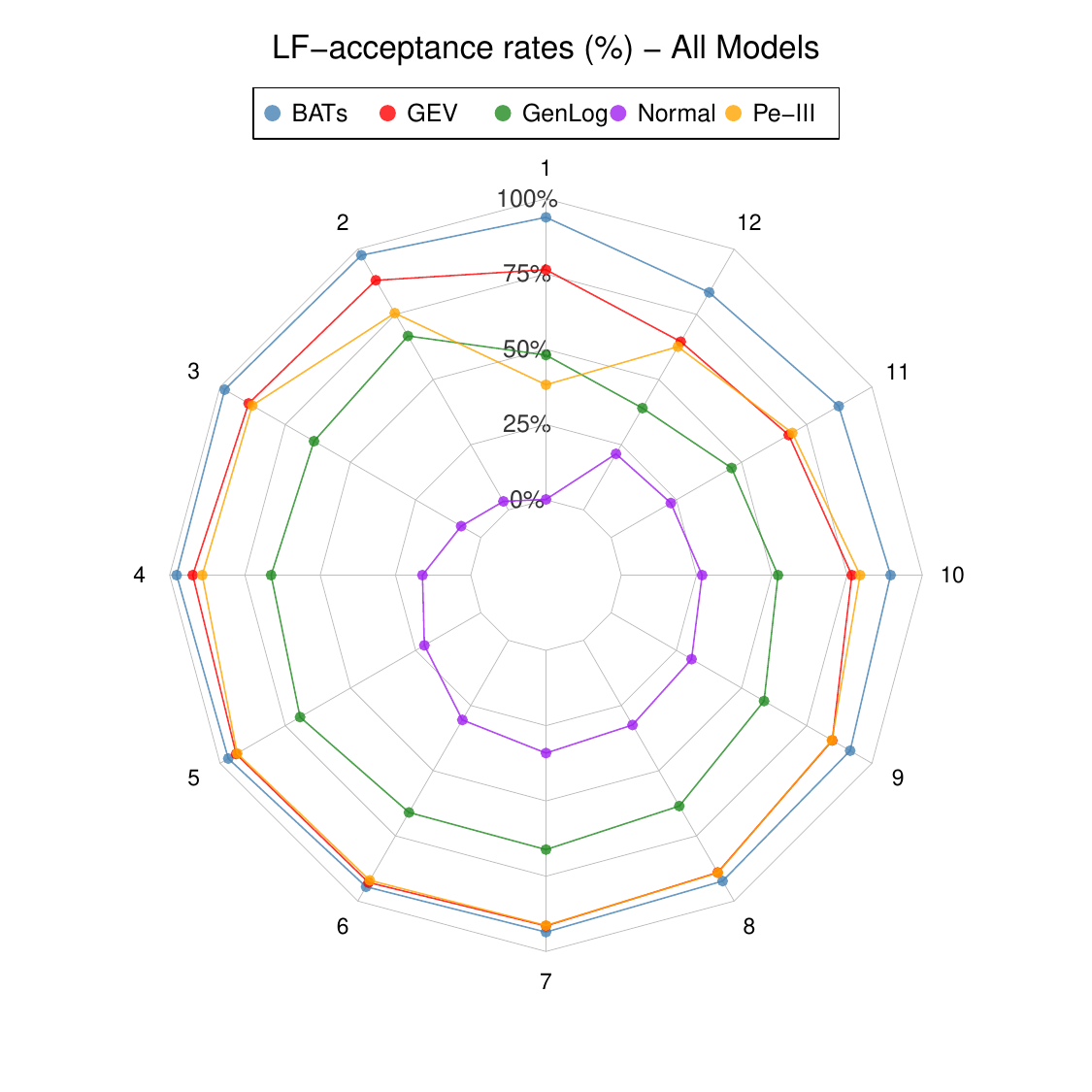}
  }
  \caption{Lilliefors normality test acceptance rate (\%) for SPEI obtained through different candidates of distributions for time scales 1 to 12. Left to right: $\mathrm{RET_{PM}}$, $\mathrm{RET_{TW}}$, and $\mathrm{RET_{HG}}$ methods were used in~$\mathcal{D}$.}
  \label{fig:LF_redar_all}
\end{figure}

In addition to the Shapiro–Wilk test, we employed further goodness-of-fit evaluations using the Anderson–Darling test and the Kolmogorov–Smirnov–based Lilliefors test (Figures~\ref{fig:AD_redar_all} and~\ref{fig:LF_redar_all}) to refine the comparison among the candidate distributions. Both tests reveal patterns that are entirely in line with the SW results and reinforce the conclusion that BATs provide the best overall fit. A comprehensive prior analysis by~\citep{stagge2015candidate} recommended the GEV distribution as a suitable option for SPEI applications in Europe. Although GEV captures extreme values reasonably well, BATs are substantially more flexible and can simultaneously adjust the behavior of both lower and upper tails.  

\begin{figure}
  \centering
  \subfigure{
\includegraphics[width=1\textwidth]{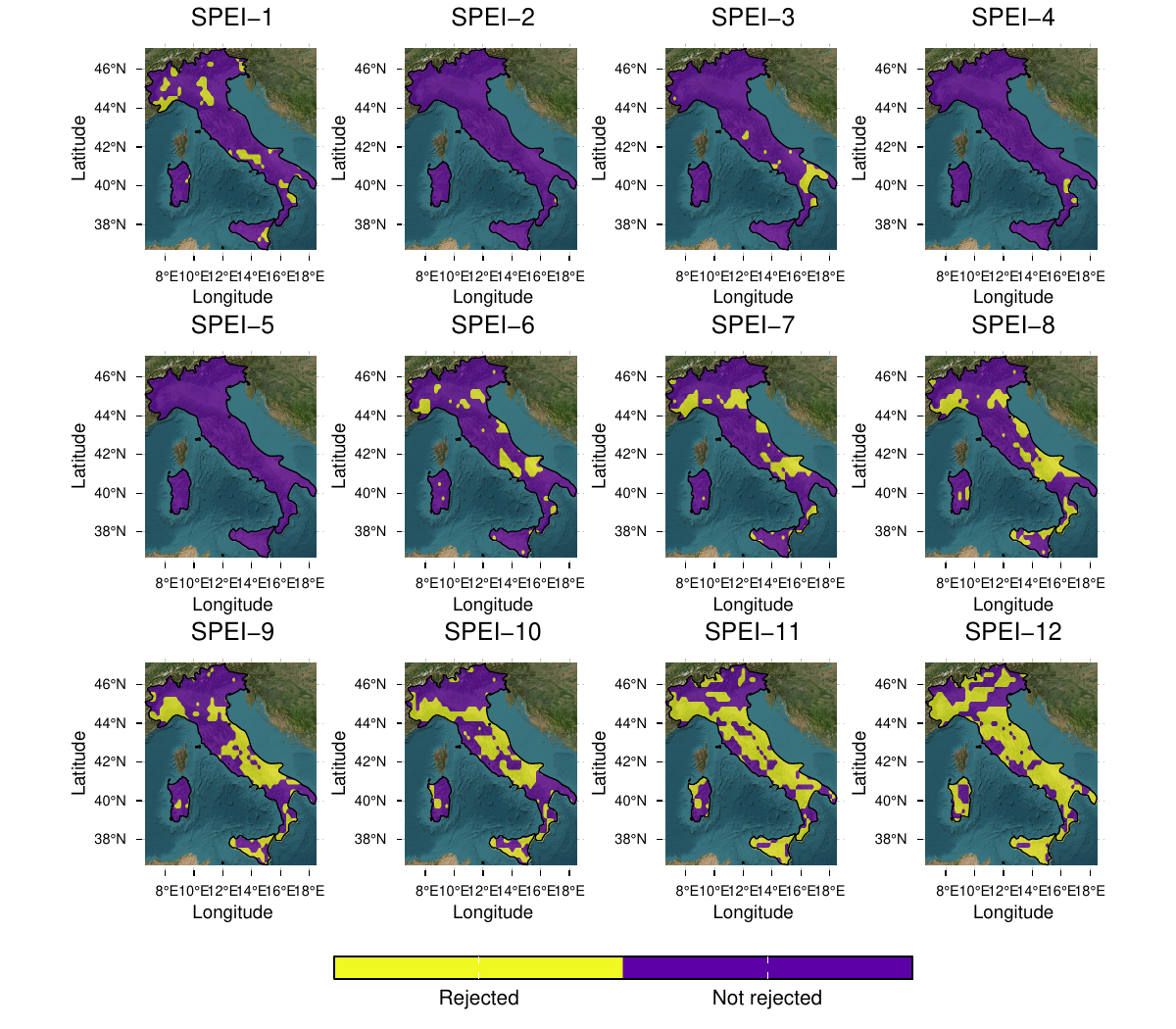}
  }
  \caption{Spatial comparison of the SW test at each grid cell under $\mathrm{RET_{PM}}$ for GEV-SPEI-1 to GEV-SPEI-12. }
  \label{fig:SW_GEV_grid}
\end{figure}

\begin{figure}
  \centering
  \subfigure{
\includegraphics[width=1\textwidth]{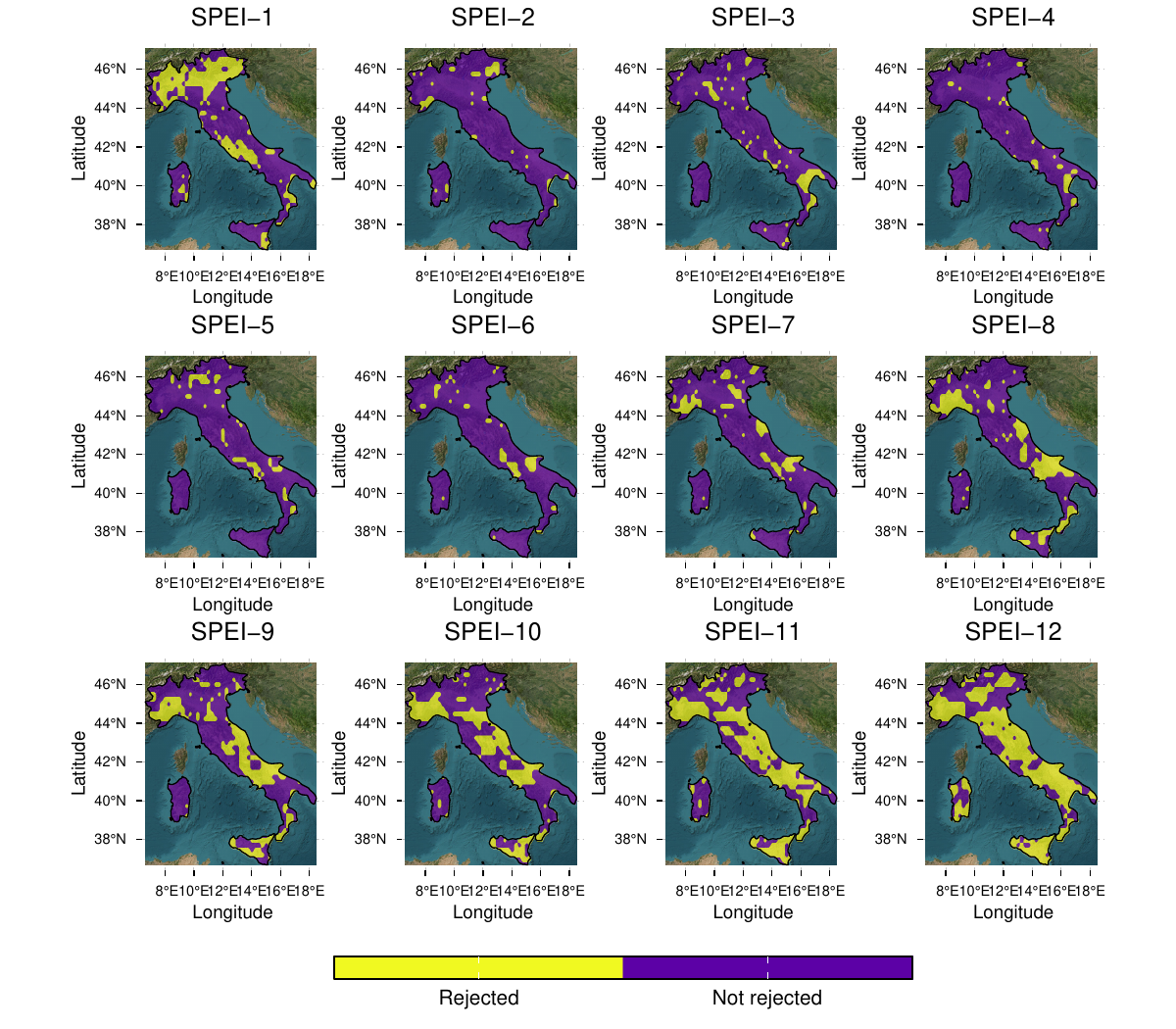}
  }
  \caption{Spatial comparison of the SW test at each grid cell under $\mathrm{RET_{PM}}$ for Pe-III-SPEI-1 to Pe-III-SPEI-12. }
  \label{fig:SW_PE3_grid}
\end{figure}

\begin{figure}
  \centering
  \subfigure{
\includegraphics[width=1\textwidth]{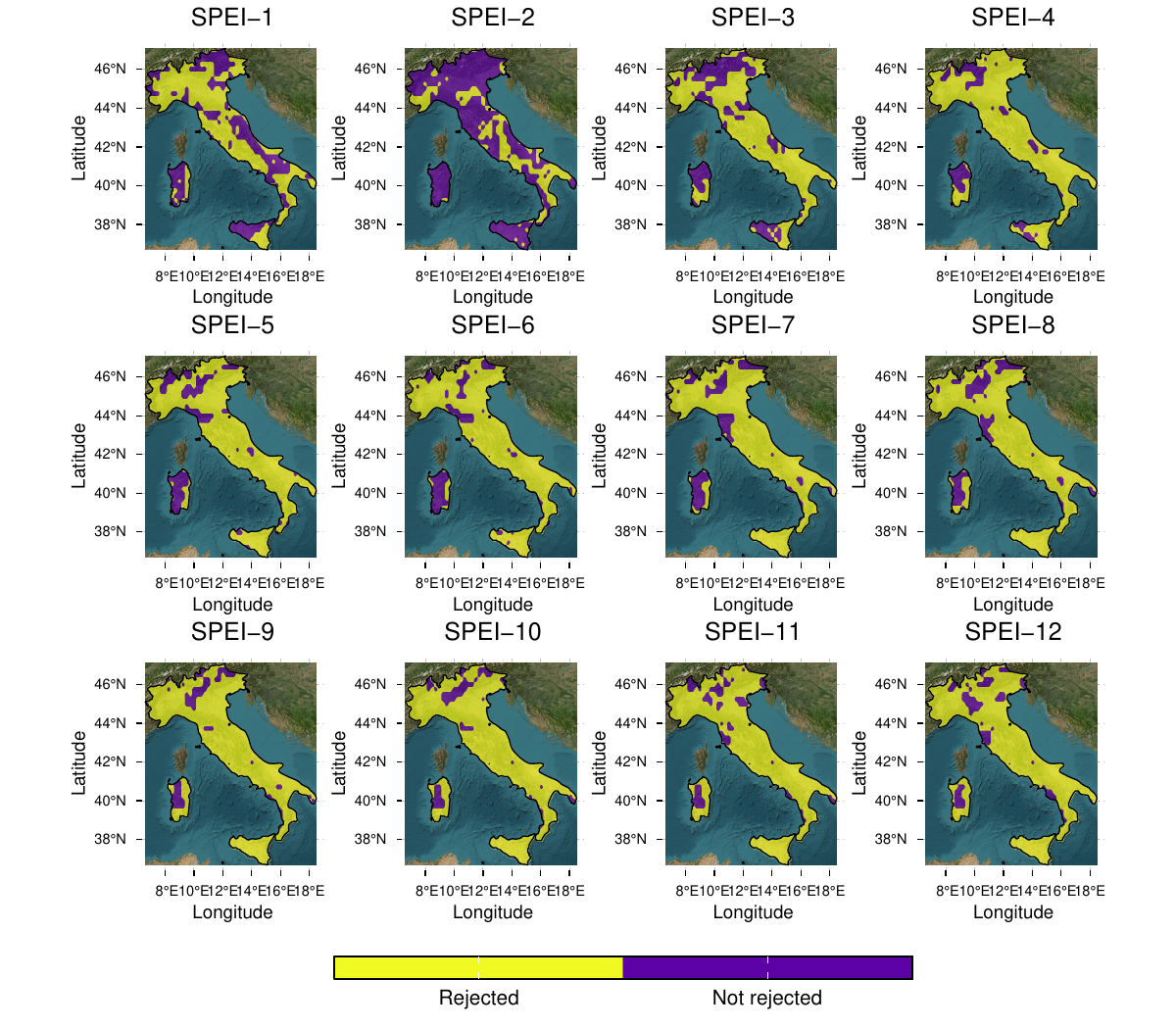}
  }
  \caption{Spatial comparison of the SW test at each grid cell under $\mathrm{RET_{PM}}$ for GenLog-SPEI-1 to GenLog-SPEI-12. }
  \label{fig:SW_Genlog_grid}
\end{figure}

\begin{figure}
  \centering
  \subfigure{
\includegraphics[width=1\textwidth]{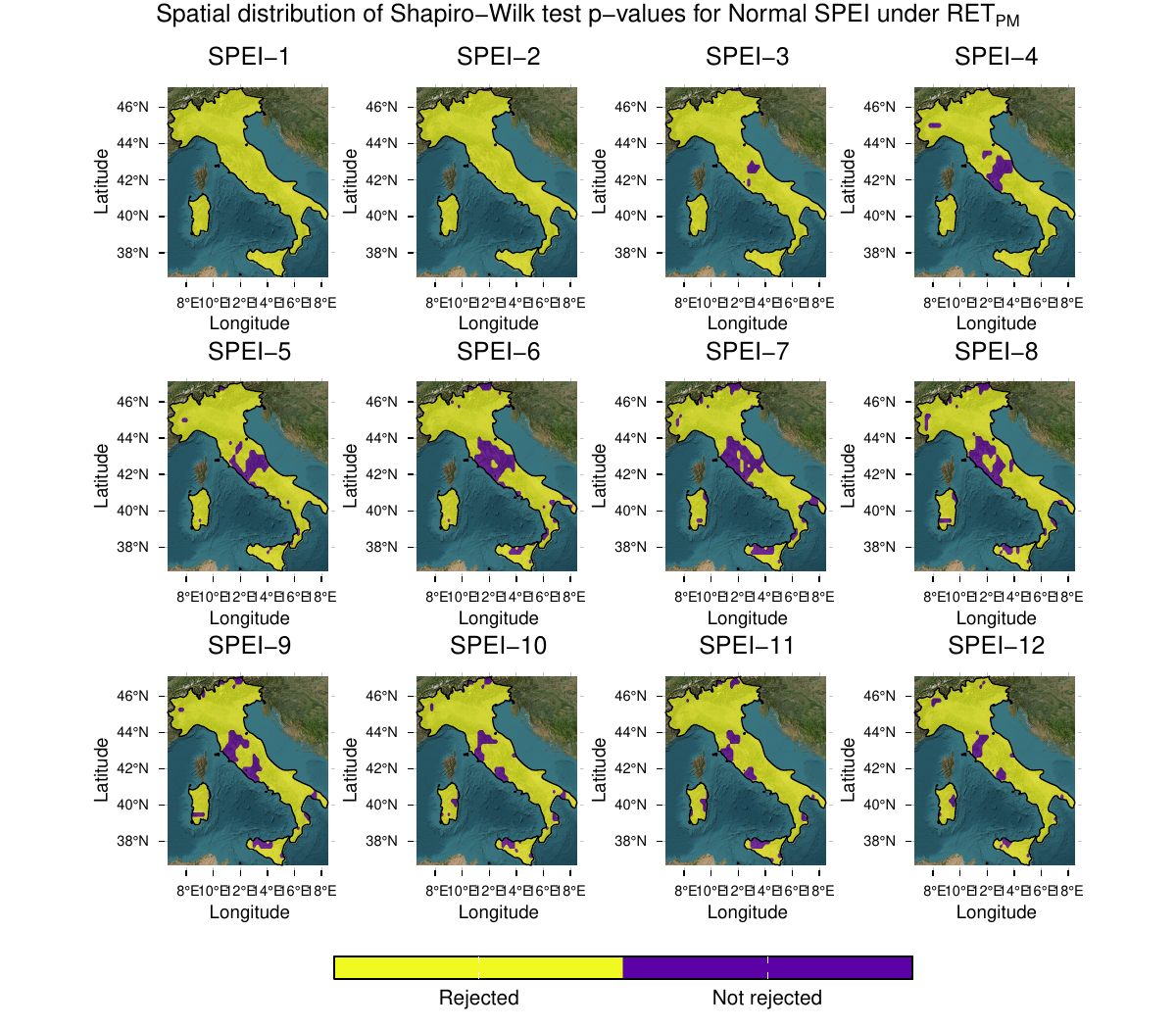}
  }
  \caption{Spatial comparison of the SW test at each grid cell under $\mathrm{RET_{PM}}$ for Normal-SPEI-1 to Normal-SPEI-12.}
  \label{fig:SW_Normal_grid}
\end{figure}

Figures~\ref{fig:SW_GEV_grid}--\ref{fig:SW_Normal_grid}
show grid-level Shapiro–Wilk test outcomes for all distributions except BATs (which are reported in the main document) at accumulation scales of 1 to 12 months under the $\mathrm{RET_{PM}}$. The grid-specific SW acceptance patterns demonstrate that BATs achieve a superior fit across a broad spectrum of climatic regimes compared to the alternative models.

\subsubsection{Spatial model for drought persistence}
\begin{figure}
  \centering
  \subfigure{
\includegraphics[width=1\textwidth]{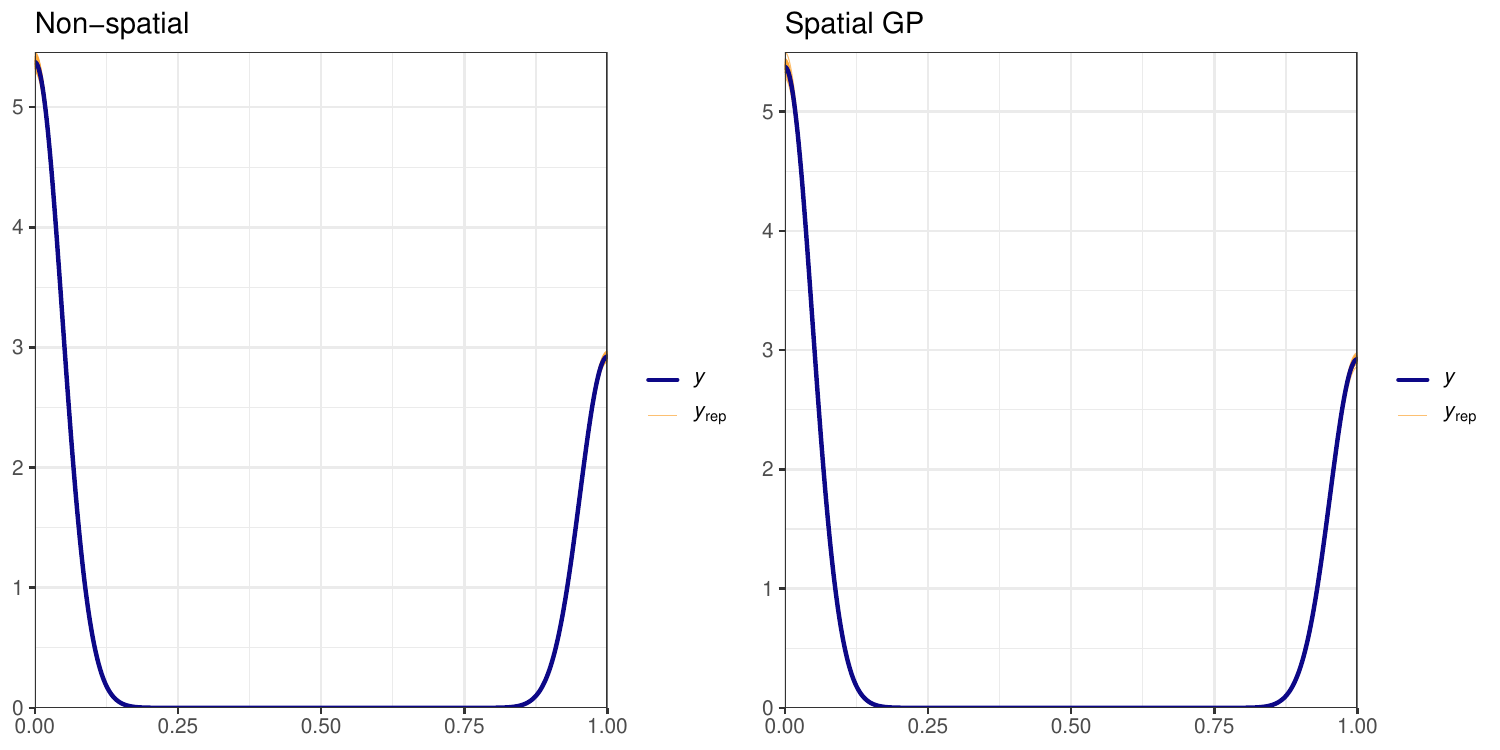}
  }
  \caption{Posterior predictive checks for nonspatial and spatial GP models.}
  \label{fig:model_diagnos}
\end{figure}

\begin{figure}
  \centering
  \subfigure{
\includegraphics[width=1\textwidth]{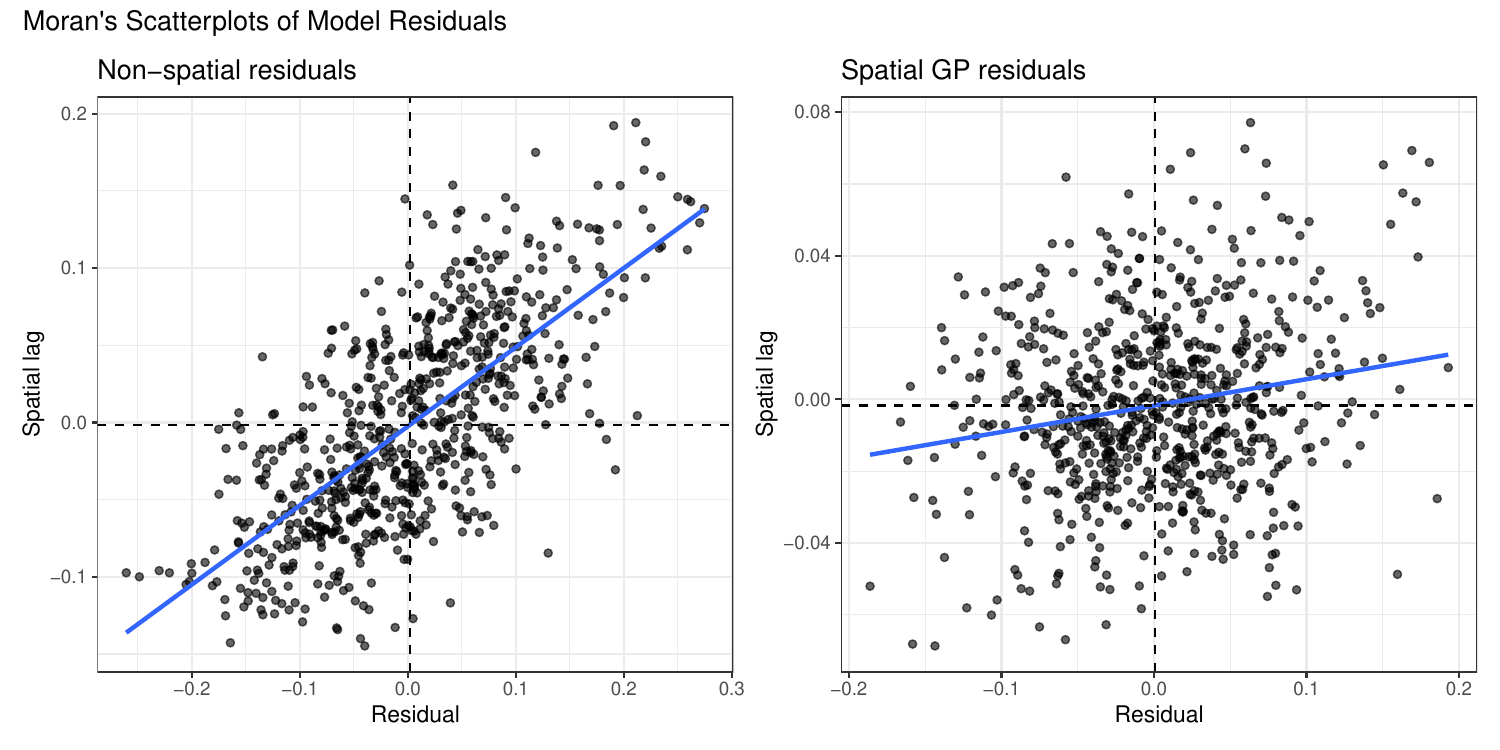}
  }
  \caption{ The Moran’ I scatterplot
of residuals of the non-spatial and spatial GP models.}
  \label{fig:model_residual}
\end{figure}

We assess model performance using posterior predictive checks and spatial diagnostics of the residuals (Figures~\ref{fig:model_diagnos}--\ref{fig:model_residual}). These checks indicate that both the non-spatial and spatial process models closely replicate the marginal distribution of the observed response, as shown by the strong agreement between observed and replicated values across the full range of the response variable. This suggests that each model adequately captures the empirical distribution of the data.

However, despite their comparable performance in reproducing the response distribution, the models differ markedly in how they handle spatial dependence. Moran's scatterplot of residuals from the non-spatial model revealed a strong positive association between residuals and their spatial lags. This substantial residual spatial autocorrelation implies that key spatial structure remains unaccounted for. By contrast, the residuals from the spatial process model exhibited no evident spatial pattern, indicating spatial independence.

 \end{document}